\pdfoutput=1
\documentclass{article}

\PassOptionsToPackage{numbers, compress}{natbib}

\usepackage[final]{neurips_2024}
\AtBeginDocument{\thispagestyle{empty}}

\usepackage[utf8]{inputenc} 
\usepackage[T1]{fontenc}    
\usepackage{url}            
\usepackage{booktabs}       
\usepackage{amsfonts}       
\usepackage{nicefrac}       
\usepackage{microtype}      
\usepackage{xcolor}         
\usepackage{enumitem}
\usepackage{graphicx}
\usepackage{xcolor}
\usepackage{colortbl}
\usepackage{array}

\usepackage[most]{tcolorbox} 
\usepackage{xcolor}          
\usepackage{soul}            
\usepackage{ragged2e}        
\usepackage{parskip}         
\usepackage{caption}         
\usepackage{geometry}        
\usepackage{xspace}
\usepackage{mdframed}
\usepackage{tikz}
\usepackage{subcaption}
\usetikzlibrary{arrows.meta, positioning, fit}
\usepackage{multirow}
\usepackage{fancyhdr}
\definecolor{hlblue}{RGB}{198,217,241}   
\definecolor{hlpurple}{RGB}{221,217,238} 
\definecolor{hlgreen}{RGB}{226,239,218}  
\definecolor{hlorange}{RGB}{255,230,204} 
\definecolor{hlteal}{RGB}{204,235,234}   

\colorlet{hlred}{red!20}
\colorlet{hlyellow}{yellow!20}
\colorlet{hlgreen}{green!20}

\newcommand{\hlYellow}[1]{\sethlcolor{hlyellow}\hl{#1}}
\newcommand{\hlGreen}[1]{\sethlcolor{hlgreen}\hl{#1}}

\newcommand{\hlSI}[1]{\sethlcolor{hlblue}\hl{#1}}     
\newcommand{\hlComp}[1]{\sethlcolor{hlorange}\hl{#1}} 
\newcommand{\hlIncl}[1]{\sethlcolor{hlgreen}\hl{#1}}  
\newcommand{\hlBiz}[1]{\sethlcolor{hlpurple}\hl{#1}}  

\newcommand{\trecrag}{TREC 2025 RAG\xspace}

\definecolor{citecolor}{RGB}{34,139,34}
\usepackage[pagebackref=true,breaklinks=true,colorlinks,
citecolor=citecolor,bookmarks=false]{hyperref}

\title{Overview of the TREC 2025 Retrieval Augmented Generation (RAG) Track}

\author{
Shivani Upadhyay$^1$, Nandan Thakur$^1$, Ronak Pradeep$^1$,\\
{\bf Nick Craswell$^2$, Daniel Campos$^3$, Jimmy Lin$^1$} \\[1ex]
$^1$ University of Waterloo \quad $^2$ Microsoft \quad $^3$ Zipf AI \\
}

\begin{document}
\maketitle

\begin{abstract}
The second edition of the TREC Retrieval Augmented Generation (RAG) Track advances research on systems that integrate retrieval and generation to address complex, real-world information needs. 
Building on the foundation of the inaugural 2024 track, this year’s challenge introduces long, multi-sentence narrative queries to better reflect the deep search task with the growing demand for reasoning-driven responses. 
Participants are tasked with designing pipelines that combine retrieval and generation while ensuring transparency and factual grounding. 
The track leverages the MS MARCO V2.1 corpus and employs a multi-layered evaluation framework encompassing relevance assessment, response completeness, attribution verification, and agreement analysis. 
By emphasizing multi-faceted narratives and attribution-rich answers from over 150 submissions this year, the TREC 2025 RAG Track aims to foster innovation in creating trustworthy, context-aware systems for retrieval augmented generation.

\begin{center}
    \textbf{Track website:} \url{https://trec-rag.github.io}
\end{center}

\end{abstract}

\vspace{2mm}
\section{Introduction}
This paper provides an overview of the TREC 2025 Retrieval Augment Generation (RAG) Track.
The \textbf{second edition} of the Retrieval Augmented Generation (RAG) Track builds on the foundation laid by the inaugural TREC 2024 RAG Track~\cite{ragnarok, trec24_ret, nugget, support}, pushing research further into systems that integrate retrieval and generation for complex, real-world information needs.
We harness the institutional knowledge and resources provided by the National Institute of Standards and Technology (NIST) via the Text Retrieval Conference (TREC) to tackle these challenges.
Now in its 34\textsuperscript{th} year, TREC has led the way in many aspects of evaluation in information retrieval (IR), natural language processing (NLP), and beyond, producing many innovations that the community (both researchers as well as practitioners) take for granted today.

Building upon last year's evaluation strategy at the TREC 2024 RAG Track, a key change this year is moving from short, keyword-style queries to multi-sentence, long and complex narratives that mimic \emph{deep search} scenarios. 
This shift reflects the growing demand for RAG and agentic systems (or agents) capable of nuanced interpretation, broader evidence coverage, and reasoning-driven responses. 
Participants are challenged to design pipelines that not only combine retrieval and generation but also maintain transparency and factual grounding in their outputs. 
These changes aim to foster innovation in creating systems that can handle intricate queries and deliver well-supported, context-rich answers.

To support these advancements, the track continues to use the MS MARCO V2.1 document corpus constructed last year at the TREC 2024 RAG Track, which contains segmentation and document deduplication, useful for assessing retrieval and generation tasks in our track. 
The corpus ensures a diverse document representation, enabling participants to build systems that can retrieve relevant evidence and generate responses grounded in verifiable and reliable sources. 
By introducing narratives and emphasizing attribution, the \trecrag Track encourages approaches that go beyond surface-level matching, requiring deeper reasoning and synthesis of information across multiple documents. 
This year, the track employs a multi-layered evaluation workflow to ensure reliability and depth:

\begin{enumerate}[leftmargin=*]
    \item \textbf{Relevance Assessment:} Human assessors judge document relevance using sub-narratives, assigning relevance scores between 0 (irrelevant) and 4 (perfectly relevant) based on the number of sub-narratives being answered in the given document.
    \item \textbf{Response Evaluation:} For relevant documents, key information units (``nuggets'') are extracted and mapped to sub-narratives to measure content coverage and answer completeness.
    \item \textbf{Support Evaluation:} Each sentence in a generated answer is checked against its cited document to verify factual grounding.
    \item \textbf{Agreement Analysis:} System labels are compared with human judgments using metrics such as Cohen’s $\kappa$ to assess alignment beyond chance.
\end{enumerate}

This layered approach combines manual and automated methods for assessment, enabling a fine-grained evaluation of retrieval quality, attribution accuracy, and response completeness. 
By emphasizing the narrative-driven queries and attribution-rich answers, the \trecrag Track aims to push the boundaries of retrieval-augmented generation research, fostering innovations that improve reasoning, coverage, and trustworthiness in AI-driven systems.

\begin{table}[t]
\setlength{\tabcolsep}{3pt}

\renewcommand{\arraystretch}{1}
\centering
\resizebox{0.95\textwidth}{!}{
{\begin{tabular}{p{16cm}}
\toprule
\toprule
{\bf Query}: I'm interested in sports' societal impact, particularly concerning athlete compensation, inclusion, cultural influence, and the business side. I also want to understand how evolving equipment, training, and mindset shape both athletes and the popularity of different sports. \\[1ex]
\midrule
{\bf Answer}: Sports exert profound societal impacts through athlete compensation, inclusion, cultural dynamics, and commercialization, while evolving equipment, training, and mindsets reshape athletic performance and sport popularity. Athlete compensation disparities persist, with female athletes and college athletes receiving inequitable pay compared to their male and professional counterparts, raising ethical questions about just compensation and the commodification of athletes \hl{[0]}. College athletes, despite generating significant revenue, often lack direct salaries, though debates persist over whether scholarships suffice or if additional compensation is warranted \hl{[0]}. Inclusion remains a challenge, as racial and gender disparities persist in leadership roles and access to opportunities. For example, Black women are underrepresented in sports like field hockey and golf, while Title IX has increased female participation but not fully addressed coaching diversity. Socioeconomic barriers further limit access to elite sports, as expensive equipment and coaching exclude lower-income individuals \hl{[2]}. Culturally, sports reflect and reinforce societal values, serving as a microcosm of racial and gender dynamics. They also foster unity, transcending social and political divides, while commercialization ties sports to consumerism, with sponsorships prioritizing brand visibility over equitable access \hl{[3]}. The business side of sports, driven by media and corporate interests, amplifies revenue but risks prioritizing profit over athlete welfare, as seen in debates over performance-enhancing drugs and gambling \hl{[6]}. Evolving equipment and training methods, such as advanced analytics and technology, enhance performance but raise ethical concerns about fairness and health risks \hl{[5]}. Mental and physical training paradigms, including discipline and resilience, influence athlete success and public perceptions of sports \hl{[4]}. Meanwhile, shifting societal values—like prioritizing health and inclusivity—reshape popular sports, with pleasure and participation sports gaining traction as alternatives to high-stakes power sports \hl{[5]}. These interwoven factors highlight sports' dual role as both a reflection of societal progress and a catalyst for change, though persistent inequities and commercial pressures underscore ongoing challenges \hl{[1]}. \\
\bottomrule
\bottomrule
\end{tabular}}}
\vspace{0.3cm}
\caption{Sample answer from the TREC 2025 RAG Track for narrative \texttt{14} with citations \hl{highlighted}.}
\vspace{-0.5cm}
\label{tab:sample-sample}
\end{table}

\section{Task Setup}
The \trecrag track comprised four interconnected tasks: (1) \emph{Retrieval} (R), (2) \emph{Augmented Generation} (AG), (3) \emph{Retrieval Augmented Generation} (RAG), and (4) \emph{Relevance Judgment} (RJ) (a new task this year); all built on the MS MARCO v2.1 segmented document collection~\cite{ragnarok}, which addressed deduplication and provided fine-grained segmentation to support robust retrieval and grounded generation. 
Participants were given 105 queries (called narratives in TREC parlance); 
their ultimate task was to return, for the AG and RAG tasks, well-formed answers for each narrative (up to a maximum of 400 words).
The Retrieval (R) task could be viewed as an intermediate product in a full RAG pipeline. Furthermore, the Relevance Judgment (RJ) task, required participants to provide a set of relevance judgments.

Throughout this paper, the narrative \texttt{14}: ``\emph{I'm interested in sports' societal impact, particularly concerning athlete compensation, inclusion, cultural influence, and the business side. I also want to understand how evolving equipment, training, and mindset shape both athletes and the popularity of different sports.}'' has been used as a running example.
A system-generated answer using Qwen3 32B~\citep{ragnarok} for the narrative is shown in Table~\ref{tab:sample-sample}.
Actual submissions to the TREC 2025 RAG Track were in the form of structured JSON, wherein each answer sentence is explicitly linked to citations of documents within the corpus that (purportedly) supported the assertions made in the sentence~\cite{ragnarok}.

Here is an overview of the four tasks conducted at the TREC 2025 RAG Track:
\begin{itemize}[leftmargin=*]
    \item \textbf{Retrieval (R):} In the Retrieval (R) task, participants are given a set of narratives (each 2--3 sentences in JSONL format with narrative id and title) and must return a TREC-style ranked list of the top-100 segment IDs from the MS MARCO V2.1 deduped segment collection~\citep{ragnarok}. Unlike previous MS MARCO data, where passages were selected to be short human-readable answers, the V2.1 passages were constructed using a sliding window approach over full documents to generate more ``natural'' content. These passages form segments (sometimes called ``chunks'') suitable for RAG.
    \item \textbf{Augmented Generation (AG):} The AG task isolates the generation component by fixing retrieval: participants receive a top‑k ranked list of segments (produced by the baseline retrieval pipeline) and must generate an answer of $\leq$ 400 words, broken into sentences, with sentence-level citations to supporting segment IDs from the provided top-k ranked list.
    \item \textbf{Retrieval Augmented Generation (RAG):} The RAG task evaluates end-to-end systems; participants bring their own retrievers, rerankers, generators, and pipeline, but must map all evidence to MS MARCO V2.1 segment IDs for evaluation. 
    \item \textbf{Relevance Judgment (RJ):} The Relevance Judgment (RJ) task determines how well individual documents address the information need expressed in a narrative. Each document is scored on a 0--4 scale, where 0 indicates the document is not relevant and 4 indicates comprehensive coverage of the narrative. This task is designed to capture varying degrees of relevance for complex, multi-sentence queries, ensuring that retrieval systems are evaluated on their ability to return documents that meaningfully contribute to answering the full narrative.
\end{itemize}

\begin{figure}[t] 
    \centering
    \includegraphics[width=\columnwidth]{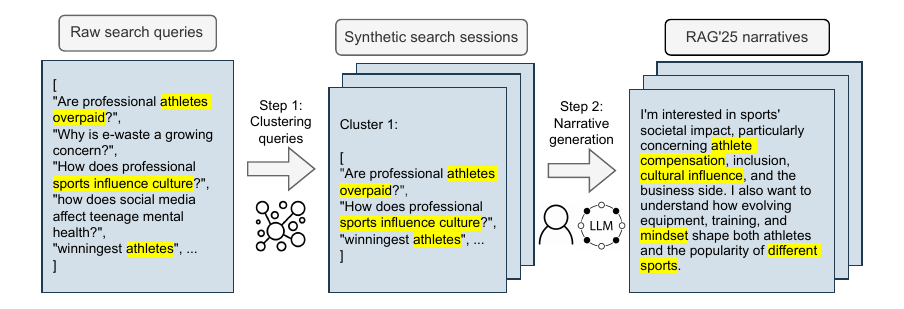}
    \caption{Overview of the narrative generation process in the TREC 2025 RAG Track.}
    \vspace{-0.3cm}
    \label{fig:overview}
\end{figure}

\paragraph{Reference Ranked Lists.}
Reference ranked lists of top-k segments were provided for those who did not wish to perform retrieval on the document collection themselves.
This represented the input to the AG task, where participants could focus on the ``augmented generation'' part of RAG.
The top documents for each narrative were provided based on an effective multi-stage ranking pipeline, described below:

The provided ranked lists included a combination of first-stage retrieval and advanced reranking strategies.
For first-stage retrieval, reciprocal rank fusion (RRF) across two high-performing models: SPLADE-v3~\cite{splade3} and Snowflake's Arctic-Embed-L~\cite{arctic} (each retrieving the top 1,000 documents) was employed.
This hybrid approach leveraged the complementary strengths of sparse and dense retrieval, as RRF has consistently demonstrated improved retrieval effectiveness across multiple evaluation benchmarks in the RAG 2024 track~\cite{ragnarok}.

Following retrieval, a reranking pipeline was deployed to refine the candidate lists.
RankLLM~\cite{rankllm}, a neural reranker was applied to process the fused retrieval results and improve narrative-document relevance.
Following this, RankQwen3-32B, a reranker operating on the top 1,000 candidates from the previous stage, produced a reordered list, from which the top 100 documents were selected.
This cascading reranking strategy ensured that the final ranked lists reflected both lexical and semantic relevance, providing participants with highly optimized results.

\paragraph{Narrative Generation.}
The \trecrag Track required narrative-style queries; however, these were not readily available, and it was cumbersome to generate them manually using human annotators. Therefore, the automated narrative generation process transformed clusters of related queries into coherent, multi-sentence narratives that simulated real-world research needs. 
For example, a cluster of queries such as ``\emph{Are professional athletes overpaid?}'' and ``\emph{How does professional sports influence culture?}'' was expanded into a narrative like: ``\emph{I’m interested in sports’ societal impact, particularly concerning athlete compensation, inclusion, cultural influence, and the business side. 
I also want to understand how evolving equipment, training, and mindset shape both athletes and the popularity of different sports.}''
Figure~\ref{fig:overview} presents an overview of the overall process.

\begin{figure}[t]
\begin{mdframed}[font=\small, roundcorner=10pt, linewidth=1pt, innerleftmargin=10pt, innerrightmargin=10pt, innertopmargin=5pt, innerbottommargin=5pt]
Generate list of atomic and exhaustive sub-queries (1-12 words) from the given narrative. Make them independent have less overlap among the sub-queries. *Try to make the sub-queries as simple as possible and non-overlapping.*\\
Each sub-query should be concise, distinct, and cover a key aspect of the narrative. Make sure there is no redundancy among the sub-queries. Also assign importance to these sub-queries. Vital sub-query represent query that must be present in a ``good'' answer for the given narrative; on the other hand, okay sub-query contribute worthwhile information about the narrative but are not essential.\\
Limit the list to at most 10.\\

Narrative: \{$narrative$\}\\

I also have a reference list of queries relevant to the narrative. Just use it as reference to generate a refine list of non-overlapping sub-queries.\\

Query List: \{$query\_list$\}\\

Return only a JSON list of dictionary object containing keys 'text' and 'importance' ('vital' or 'okay' w.r.t narrative), no extra commentary.
\end{mdframed}
\caption{Prompt used in the first step of the sub-narrative generation process with GPT-4.1 nano to generate an exhaustive list of sub-queries using the given narrative and query cluster as input.}
\vspace{-0.1cm}
\label{fig:prompt_1_iter}
\end{figure}
\begin{figure}[t]
\begin{mdframed}[font=\small, roundcorner=10pt, linewidth=1pt, innerleftmargin=10pt, innerrightmargin=10pt, innertopmargin=5pt, innerbottommargin=5pt]
Re-write list of atomic and exhaustive sub-queries (1-12 words) from the given narrative. Make them independent have less overlap among the sub-queries. *Try to make the sub-queries as simple as possible and non-overlapping.*\\
Each sub-query should be concise, distinct, and cover a key aspect of the narrative. Make sure there is no redundancy among the sub-queries. Also assign importance to these sub-queries. Vital sub-query represent query that must be present in a “good” answer for the given narrative; on the other hand, okay sub-query contribute worthwhile information about the narrative but are not essential.\\
Limit the list to at most 10.\\

Narrative: \{$narrative$\}\\

I am providing list of queries for the narrative. Just use it ans reference to re-write a refine and *non-redundant* list of queries.\\

Query List: \{$previous$ $sub\_queries$\}\\

Return only a JSON list of dictionary object containing keys 'text' and 'importance' ('vital' or 'okay' w.r.t narrative), no extra commentary.
\end{mdframed}
\caption{Prompt used in the second step of refining the sub-narratives with GPT-4.1 to generate an exhaustive list of sub-queries using the given narrative and the sub-queries generated in the previous step as input.}
\vspace{-0.3cm}
\label{fig:prompt_2_iter}
\end{figure}
\begin{figure}[t]
\begin{mdframed}[font=\small, roundcorner=10pt, linewidth=1pt, innerleftmargin=10pt, innerrightmargin=10pt, innertopmargin=5pt, innerbottommargin=5pt]
\textbf{Narrative:} I'm interested in sports' societal impact, particularly concerning athlete compensation, inclusion, cultural influence, and the business side. I also want to understand how evolving equipment, training, and mindset shape both athletes and the popularity of different sports. \\

\textbf{Sub-narratives:} \\
Societal impact of sports \\
Athlete compensation issues \\
Business influences in sports \\
Sports and cultural influence \\
Inclusion in sports \\
Advancements in sports equipment \\
Innovations in athlete training \\
Role of athlete mindset changes \\
How equipment affects sport popularity 
\end{mdframed}
\caption{Example of narrative and sub-narrative pair used in \trecrag.}
\label{fig:sample}
\vspace{-0.3cm}
\end{figure}

\paragraph{Sub-Narratives Generation.}
For each narrative, a decomposition was undertaken to identify and construct a comprehensive set of sub-narratives that collectively capture all dimensions of the original narrative. 
These sub-narratives were derived from both the initial query cluster to generate the primary narrative and the narrative text itself. The overall process for sub-narrative generation was semi-automated.

The first phase of this process utilized GPT‑4.1 nano, which was provided with the narrative and its corresponding query list (used to generate the narrative) and was instructed to generate an exhaustive collection of candidate sub-narratives (see Figure~\ref{fig:prompt_1_iter}). 
Each prompt produced a JSON-formatted list of dictionary objects, where each object included a text field describing the sub-narrative and an importance field denoting its relevance to the narrative (vital or okay).
In the second phase, both the original narrative and the preliminary sub-narrative list obtained from the preceding step are supplied to GPT‑4.1, which generated a refined and more coherent set of sub-narratives while adhering to the same JSON output structure (see Figure~\ref{fig:prompt_2_iter}).
Finally, each narrative and sub-narrative pair was systematically reviewed and analyzed by human annotators. 
When necessary, the resulting sub-narrative lists were subsequently post-edited to ensure completeness, coherence, and alignment with the intended narrative scope.
Figure~\ref{fig:sample} illustrates an example of the narrative and sub-narrative list generated using this process.

\section{Evaluation}

\subsection{Relevance Assessment}
Relevance assessment at the \trecrag Track focuses on determining how well retrieved documents satisfy complex, multi-faceted information needs expressed through narratives. 
These narratives often encompass diverse aspects, making direct judgment on the entire narrative challenging.
To address this complexity, the process begins by decomposing each narrative into a set of sub-narratives, which represent simpler and more focused components of the original information need. 
This decomposition enables assessors to evaluate relevance in a structured and consistent manner, reducing ambiguity and improving reliability across judgments.
\begin{figure*}[t]
\begin{mdframed}[font=\small, roundcorner=10pt, linewidth=1pt, innerleftmargin=10pt, innerrightmargin=10pt, innertopmargin=5pt, innerbottommargin=5pt]
Given a user first person narrative and a passage, you must provide a score on an integer scale of 0 to 4 with the following meaning:\\

0 - represents that the passage has nothing to do with the narrative,\\
1 - represents that the passage seems related to the query but does not contain any sub-narrative of an answer to it,\\
2 - represents that the passage contains detailed answer for the 1 sub-narrative of the narrative with enough description,\\
3 - represents that the passage contains detailed answer for the 2 to 3 sub-narratives of the narrative with enough description and\\
4 - represents that the passage contains detailed answer for the 4+ sub-narratives of the narrative with enough description.\\

Narrative: \{$narrative$\}\\
Sub-narratives: \{$sub\_narratives$\}\\
Passage: \{$passage$\}\\

Instructions:
Determine whether each sub-narrative is fully and properly explained or only briefly mentioned.
Assign a final integer score (0-4) based on the number of sub-narratives addressed.

Rule: A passage must cover a sub-narrative in a detailed, proper way to count; mere mentions or vague references do not qualify. If there is some extra information not relevant to the narrative at all, downgrade the level by 1. If lot of extra information then downgrade by 2.
Do not provide reasoning after listing answered parts; only give the score in the format: \#\#final score: X.
\end{mdframed}
\caption{Prompt for assigning relevance labels, where the inputs are the sub-narrative list, narrative and passage pair.}
\label{fig:prompt_w_sub_narr}
\end{figure*}

\begin{table}[t]
\centering
\renewcommand{\arraystretch}{1.2}
\begin{tabular}{p{6.5cm} p{6.5cm}}
\toprule
\textbf{Document A} & \textbf{Document B} \\
\midrule
\tiny
Experience clearly shows that sport is one of the most effective mean of re-education, providing opportunities for the realization of energy, initiative, creativity and physical abilities, for leadership and competition in a healthy manner. Some of the problems of sport in modern society ``We are glad tolive in the times of the Olympic movement, which aims participation in building a peaceful and better world, through education of youth through sport, in a spirit of Olympism and without any discrimination. This requires us to a friendly understanding, solidarity and fair play'' Juan Antonio Samaranch. \hlSI{We can say that sport is a mirror which, in accordance with its objectives and the activity, reflects the basic values of society, without exerting an active influence on them. The historical process of organized sports movement about a century shows that in Western countries prerequisites for fast development of sport are based on contemporary social changes. The very emergence of high competitiveness in the sport was a reflection of a competitive society, associated with liberal thinking. A national physical culture reflects the basic trends of the values of society, and international sporting life is a barometer of international politics.} Find out how UKEssays.com can help you! Our academic experts are ready and waiting to assist with any writing project you may have. From simple essay plans, through to full dissertations, you can guarantee we have a service perfectly matched to your needs.
\newline
\newline
\textbf{Key Coverage:} \hlSI{Societal impact}.
\newline
\textbf{Label:} \colorbox{yellow!20}{\textbf{2 (Relevant)}} 
&
\tiny
\hlSI{College Scholarship Opportunities The growth of sports has also made possible the education of many young people who would otherwise never have that opportunity.} Though today most of the noticeable college athletes often leave school early to enter the professional ranks of their sport, the vast majority of college athletes graduate and move on to a field other than their sport. [...] \hlIncl{Integration We will examine how sports embraced (or at least shoe-horned) integration, looking at Willie O'Ree playing hockey for the Boston Bruins, Prentiss Gault, a football player at the University of Oklahoma, Kenny Washington, the first} \hlComp{African-American to play in the NFL, the whole-scale, and early, integration by the NBA, soccer's struggle to this day with racism, particularly in Europe, and, perhaps most famously, Jackie Robinson's entrance into baseball.} Big Business It is also impossible to talk about modern sports without considering the heavy influence of business. \hlBiz{Sports and business have become forever linked, even in college, which is supposedly played by ``amateurs'' and not ``professionals,'' despite the fact that the coaches and athletic directors make as much as their counterparts in the professional leagues.} From sports on television to sponsorships to naming rights on stadiums, the history of the business of sports reveals that business tied itself to sports more and more in the latter half of the 20th century, really ramping up in the 1970's and onward. [...]
\newline
\newline
\textbf{Key Coverage:} \hlIncl{Popularity of different sports}, \hlBiz{Business influence}, \hlComp{Athlete compensation}, \hlSI{Societal impact}.
\newline
\textbf{Label:} \colorbox{green!20}{\textbf{4 (Perfectly Relevant)}} \\
\bottomrule
\end{tabular}
\vspace{0.3cm}
\caption{Relevance assessment for narrative \texttt{14}: \textit{``I'm interested in sports' societal impact, particularly concerning athlete compensation, inclusion, cultural influence, and the business side. I also want to understand how evolving equipment, training, and mindset shape both athletes and the popularity of different sports.''}}
\vspace{-0.5cm}
\label{tab:rel-assess}
\end{table}

The original narrative is broken down into distinct sub-narratives that capture specific thematic dimensions. 
For example, the broad narrative about sports and its societal impact is decomposed into focused areas such as \emph{athlete compensation issues}, \emph{sports and cultural influence}, \emph{role of athlete mindset changes}, \emph{societal impact of sports}, and \emph{business influences in sports}. 
These sub-narratives make it easier to analyze and address different aspects of the overarching narrative. 
This structured approach ensures that complex narratives are organized into manageable, well-defined components for deeper exploration.

Once sub-narratives are established, assessors assign relevance labels to segments using a five-point scale based on coverage: 0 (\emph{Irrelevant}) for segments unrelated to the narrative, 1 (\emph{Related}) for segments offering general but non-specific information, 2 (\emph{Relevant}) for content addressing one sub-narrative, 3 (\emph{Highly Relevant}) for segments covering two or three sub-narratives, and 4 (\emph{Perfectly Relevant}) for those addressing four or more sub-narratives. 
This approach simplifies the evaluation of complex narratives, ensures alignment between assessors, and provides a robust foundation for subsequent stages such as nugget extraction and attribution. 
By focusing on sub-narrative coverage rather than the entire narrative, the relevance assessment process enhances interpretability and fairness in evaluating retrieval and generation systems~\cite{freshstack}.
Table~\ref{tab:rel-assess} compares the relevance assessment labels for two documents, where Document A focuses solely on the societal aspect of the narratives and is therefore assigned a label of 2. 
In contrast, Document B addresses approximately four different aspects of the narratives, and is assigned a label of 4.

\paragraph{Automatic Assessment.}
\label{sec:rel_auto}
To automate relevance assessments, an approach inspired by UMBRELA~\cite{umbrela} is employed. 
To ensure diversity and robustness in judgments, an ensemble of cutting-edge large language models (LLMs), including GPT‑4.1, Gemini 2.5 Pro, Qwen3 Thinking 32B, and GPT‑OSS (20B/120B), is utilized to generate and cross-validate relevance scores.
Figure~\ref{fig:prompt_w_sub_narr} showcases the exact prompt used for performing individual assessments.
The models were provided with the list of sub-narratives following a similar approach to that used by NIST assessors for manual assessment.

\subsection{Response Evaluation}
Response evaluation in \trecrag follows a two-step process to measure both the quality and coverage of system outputs.

\begin{itemize}[leftmargin=*]
    \item \textbf{Step 1: Nugget Generation and Assignment} involves identifying and classifying key information units (nuggets) from relevant documents and determining their presence in system responses.
    \item \textbf{Step 2: Sub-Narrative Mapping} aligns these nuggets with sub-narratives to assess how comprehensively the response addresses the intended information need. Table~\ref{tab:nugget-subnarrative} (left) illustrates the nugget to sub-narrative example for the sports narrative.
\end{itemize}

For the assignment, strict vital recall is used, as it is found to be best suited for response evaluation in the TREC RAG 2024 Track~\cite{nugget}.
In particular, this metric calculates recall over all vital nuggets that are fully supported by the response.
Table~\ref{tab:nugget-subnarrative} (right) presents the assignment for the sports example.
The sub-narrative \emph{Athlete compensation issues} is marked as covered since at least one of the mapped nuggets is fully supported by the response used for this example.

\paragraph{AutoNuggets + AutoAssign.}
\label{sec:auto_setting}

\begin{table}[t]
\centering
\resizebox{\columnwidth}{!}{%
\begin{tabular}{p{8cm}@{\quad} @{\quad}c@{\quad} @{\quad}l ||  @{\quad}c@{\quad} @{\quad}c@{\quad}}
\toprule
\multirow{2}{*}{\textbf{Nugget}} & & \textbf{Sub-Narrative} & \textbf{Nugget} & {\textbf{Sub-Narrative}} \\
 & & \textbf{Mapping} & \textbf{Assignment} & {\textbf{Coverage}} \\
\midrule
\small Athlete compensation is highly unequal and controversial (vital) & $\rightarrow$
& \small Athlete compensation issues &  Full Support & Yes \\

\small International events such as the Olympics and World Cup foster national pride (vital) & $\rightarrow$
& \small Sports and cultural influence & Partial Support & No \\

\small Business interests shape athlete pay, media rights, and management structures (vital) & $\rightarrow$
& \small Business influences in sports & Partial Support & No \\

\small Sports foster social cohesion and community identity & $\rightarrow$
& \small Societal impact of sports & Full Support & Yes \\

\small Major leagues profit most from broadcast deals, sponsorships, and merchandising & $\rightarrow$
& \small Business influences in sports & No Support & No \\

\bottomrule
\end{tabular}}
\vspace{0.3cm}
\caption{The left side of the table illustrates the mappings between nuggets and sub-narratives for narrative \texttt{14}, while the right side demonstrates how these assignments are carried out for metric calculation based on a chosen response.}
\vspace{-0.3cm}
\label{tab:nugget-subnarrative}
\end{table}

In this setting, both steps are fully automated by LLMs. 
Nugget generation begins by extracting concise, semantically meaningful information units from all documents marked as related or higher (relevance label $\geq$ 1). 
This is implemented using AutoNuggetizer~\cite{nugget}, a system that integrates multiple LLMs, including Gemini 2.5 Pro, GPT-4.1, Qwen3 Thinking 32B, and GPT-OSS 120B. 
Each model independently produces candidate nuggets, which are then merged at last using GPT-4.1 to ensure consistency and eliminate redundancy.

AutoNuggetizer also classifies nuggets into two categories: vital and okay. 
A vital nugget represents indispensable information required to fully answer the narrative, while an okay nugget adds helpful but non-essential details.
Using these nuggets, AutoNuggetizer assigns nugget-level support labels to system responses.
AutoNuggetizer adopts a listwise approach for nugget assignment, where the LLM (here GPT-4.1) analyzes an answer and determines whether each nugget is fully supported (Full Support), partially supported (Partial Support), or not supported (No Support) by the answer.

The second step, sub-narrative mapping, uses GPT-4.1 to align each nugget with its corresponding sub-narrative. 
If a nugget does not fit any existing sub-narrative, a new sub-narrative is created to maintain comprehensive coverage. 
Coverage is calculated as follows: a sub-narrative is considered covered if at least one mapped nugget is fully supported by the response.

\paragraph{Post-Edit Nuggets + AutoAssign.}
In this setting, nuggets are generated automatically using the same model combination (same as AutoNuggets), and are further post-edited manually by NIST assessors. 
The post-editing process may involve rewriting nuggets for clarity, breaking down complex nuggets into smaller units, and adjusting their importance to better reflect relevance.

Additionally, NIST assessors perform sub-narrative mapping during this stage. 
Each nugget is mapped to the most relevant sub-narrative, and if a group of nuggets does not fit any existing sub-narrative, a new one is created to ensure comprehensive coverage.

Finally, nugget assignment to system responses is carried out using AutoNuggetizer.
For sub-narrative coverage, the manual mappings are combined with the automated assignments to calculate the number of sub-narratives covered in a given response.

\paragraph{Post-Edit Nuggets + ManualAssign.}
This represents the most manual setting for response evaluation.
In this setting, post-edited nuggets are used, and NIST assessors perform the assignment step.
Assessors examine each response and assign a support level (Full Support, Partial Support or No Support) based on their judgment.

\begin{table}[t]
\centering
\renewcommand{\arraystretch}{1.2}
\begin{tabular}{p{8cm} p{5cm}}
\toprule
\textbf{Document} & \textbf{Answer Sentence} \\
\midrule
\tiny
College Scholarship Opportunities \hlGreen{The growth of sports has also made possible the education of many young people who would otherwise never have that opportunity.} Though today most of the noticeable college athletes often leave school early to enter the professional ranks of their sport, the vast majority of college athletes graduate and move on to a field other than their sport. \hlGreen{For many of those students, sports made going to college possible, particularly through the use of college scholarships, which have increased in size and number as sports created bigger revenues for schools.} Integration We will examine how sports embraced (or at least shoe-horned) integration, looking at Willie O'Ree playing hockey for the Boston Bruins, Prentiss Gault, a football player at the University of Oklahoma, Kenny Washington, the first African-American to play in the NFL, the whole-scale, and early, integration by the NBA, \hlYellow{soccer's struggle to this day with racism, particularly in Europe, and, perhaps most famously, Jackie Robinson's entrance into baseball}. \hlGreen{Big Business It is also impossible to talk about modern sports without considering the heavy influence of business.} Sports and business have become forever linked, even in college, which is supposedly played by ``amateurs'' and not ``professionals,'' despite the fact that the coaches and athletic directors make as much as their counterparts [...]
&
\tiny
\textbf{Sentence:} COVID-19 disrupted training, sponsorships, and sports education but also spurred sector-wide solidarity and public-health messaging across the value chain.
\newline
\textbf{Label:} \colorbox{red!20}{\textbf{No Support}} 
\newline
\newline
\textbf{Sentence:} Landmark events such as Jackie Robinson breaking racial barriers continue to resonate, though racial inequality and representation remain critical concerns.
\newline
\textbf{Label:} \colorbox{yellow!20}{\textbf{Partial Support}} 
\newline
\newline
\textbf{Sentence:} The societal impact of sports is multifaceted, encompassing athlete compensation, inclusion, cultural influence, and the business side of sports.
\newline
\textbf{Label:} \colorbox{green!20}{\textbf{Full Support}} \\
\bottomrule
\end{tabular}
\vspace{0.3cm}
\caption{Support assessment example for narrative \texttt{14}. The relevant parts of the document are highlighted that provide information to support information in the answer sentence.}
\vspace{-0.5cm}
\label{tab:support-assess}
\end{table}
\subsection{Support Evaluation}
Support evaluation represents an attempt to determine whether information in a RAG answer is factually supported by its cited document~\cite{support}.
Evaluation in the \trecrag is conducted by breaking down the answer, with the unit of evaluation defined at a \textbf{sentence-level} (unlike response evaluation).
An answer $r$ is assumed to be segmented into $n$ sentences, $r = \{a_1, \cdots, a_n\}$, where each sentence $a_i$ can contain a maximum of $k$ document citations $[d_1,\cdots, d_k]$, with each document drawn from a corpus.
Support is calculated as the function $f(a_i, d_j)$ where $f$ can be a human or LLM assessor that generates a scalar value $s_{i,j}$, indicating the extent to which the cited document $d_j$ provides support for sentence $a_i$.

As with all
TREC evaluations, NIST provides the resources to perform human evaluations based on the guidance of the track organizers. 
NIST first trains every human assessor to understand the task, and then each human assessor evaluates each narrative sequentially. 
This year, the support evaluation focuses on sparse annotations by restricting to evaluate only the first (or highest ranked) cited document, similar to last year~\cite{support}. 
For evaluation, NIST assessors use a three-level grade, with the following descriptions:

\begin{enumerate}[leftmargin=*]
    \item \textbf{Full Support}: All of the information in the answer sentence is factually consistent with and supported by the cited segment.
    \item \textbf{Partial Support}: Some of the information in the answer sentence is factually consistent with and supported by the cited segment, but other parts of the sentence are not supported.
    \item \textbf{No Support}: The cited segment is completely irrelevant and does not support any part of the answer sentence.
\end{enumerate}

An edge case arises when a sentence contains zero citations: the support level is automatically considered ``No support'', as the sentence does not cite any retrieved segment. This year, two cases are considered for support evaluation: \textbf{Automatic}, in which GPT-OSS 120B performs the automatic assessment, and \textbf{Manual}, in which NIST assessors perform the evaluation and assign a support level to each answer--document pair (Full Support, Partial Support or No Support) based on their judgment.

Support is evaluated across two dimensions, following Liu et al.~\cite{liu-etal-2023-evaluating}: (1) weighted precision, which accounts for how many correct passage citations are present in the generated answer, and (2) weighted recall, which accounts for how many sentences in the answer are supported by passage citations. Both metrics are defined below:

\paragraph{Weighted Precision.} This metric measures the weighted proportion of citations that support each answer sentence.
For $s(a_i, d_j)$, a weight of $1.0$ to Full Support (FS), $0.5$ to Partial Support (PS), and $0$ to No Support (NS) is assigned for the answer sentence and cited passage. 
To explain the metric clearly, consider a RAG answer with $3$ sentences = $\{a_1, a_2, a_3\}$, and a corpus $C$ with $2$ passages:\ $\{p_1, p_2\}$. 
Suppose, passage $p_1$ partially supports $a_1$, passage $p_2$ fully supports $a_2$, and $a_3$ has zero citations. 
Weighted precision is computed as follows:
\begin{equation*}
\text{Weighted Precision} = \dfrac{s(a_1,p_1) + s(a_2,p_2)}{\text{count}(\{a_1,p_1\},\{a_2,p_2\})} = \dfrac{0.5 + 1}{2} = 0.75
\end{equation*}

\paragraph{Weighted Recall.} 
This metric measures the weighted proportion of answer sentences that are supported by their cited passages. 
The same weights defined above in weighted precision are applied.
For the above example, weighted recall is computed as follows:

\begin{equation*}
\text{Weighted recall} = \dfrac{s(a_1,p_1) + s(a_2,p_2)}{\text{count}(\{a_1, a_2, a_3\})} = \dfrac{0.5 + 1}{3} = 0.5
\end{equation*}

\subsection{Relevance Judgment Evaluation}
To assess the reliability of automated labeling, the pairwise agreement, ties, and disagreements between human judgments and participants' labels are evaluated. 
This comparison helps identify where automated methods align with expert assessments and where discrepancies occur. 
For runs in which the automated labels fall within the same scale as human labels (0--4), Cohen’s Kappa score is computed to measure the strength of agreement beyond chance. 
This metric provides a quantitative indicator of alignment quality, ensuring that the evaluation captures both consistency and potential gaps between human and machine judgments.

 \begin{figure*}[t]
    \includegraphics[width=0.32\textwidth]{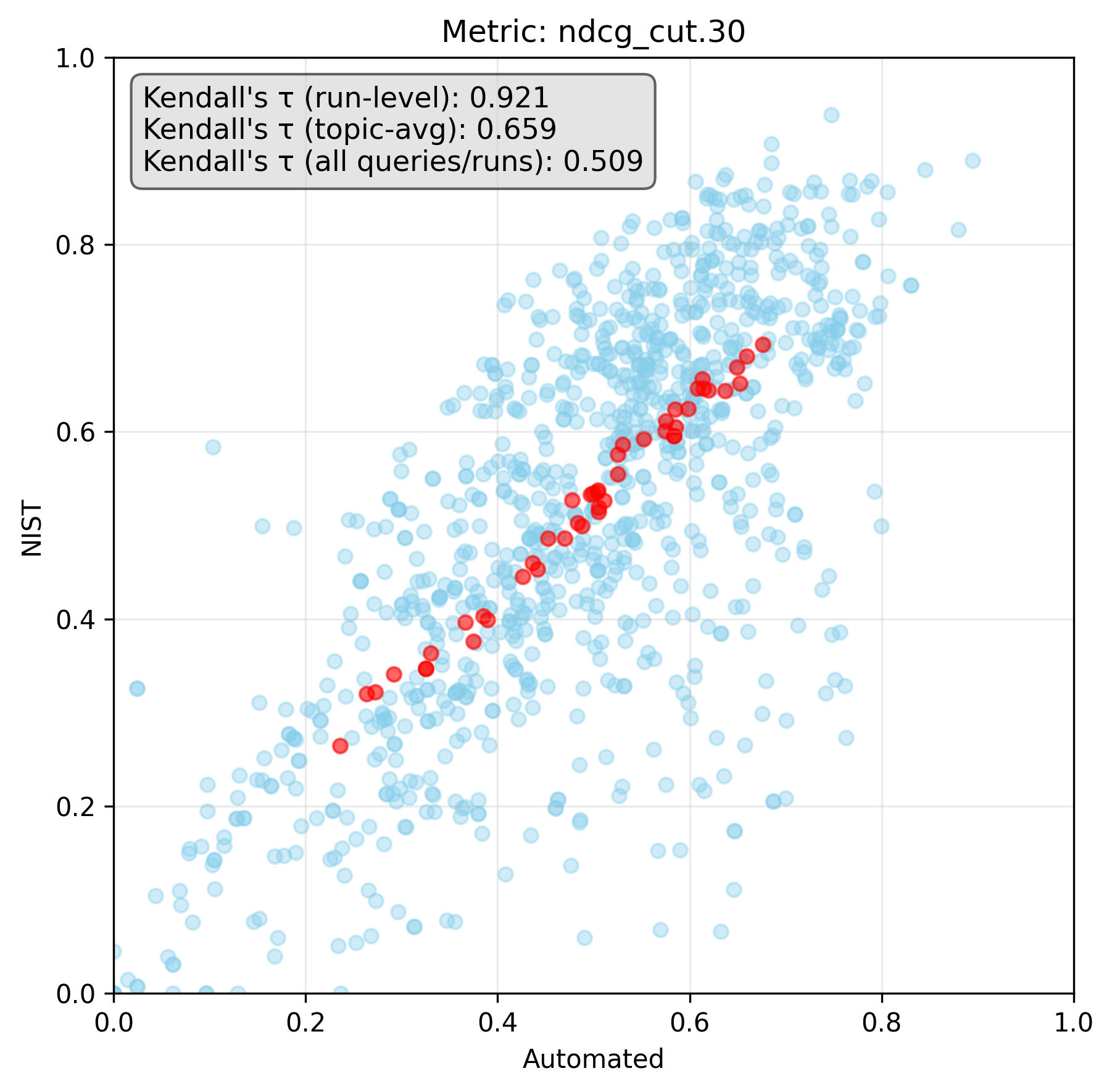}
    \includegraphics[width=0.32\textwidth]{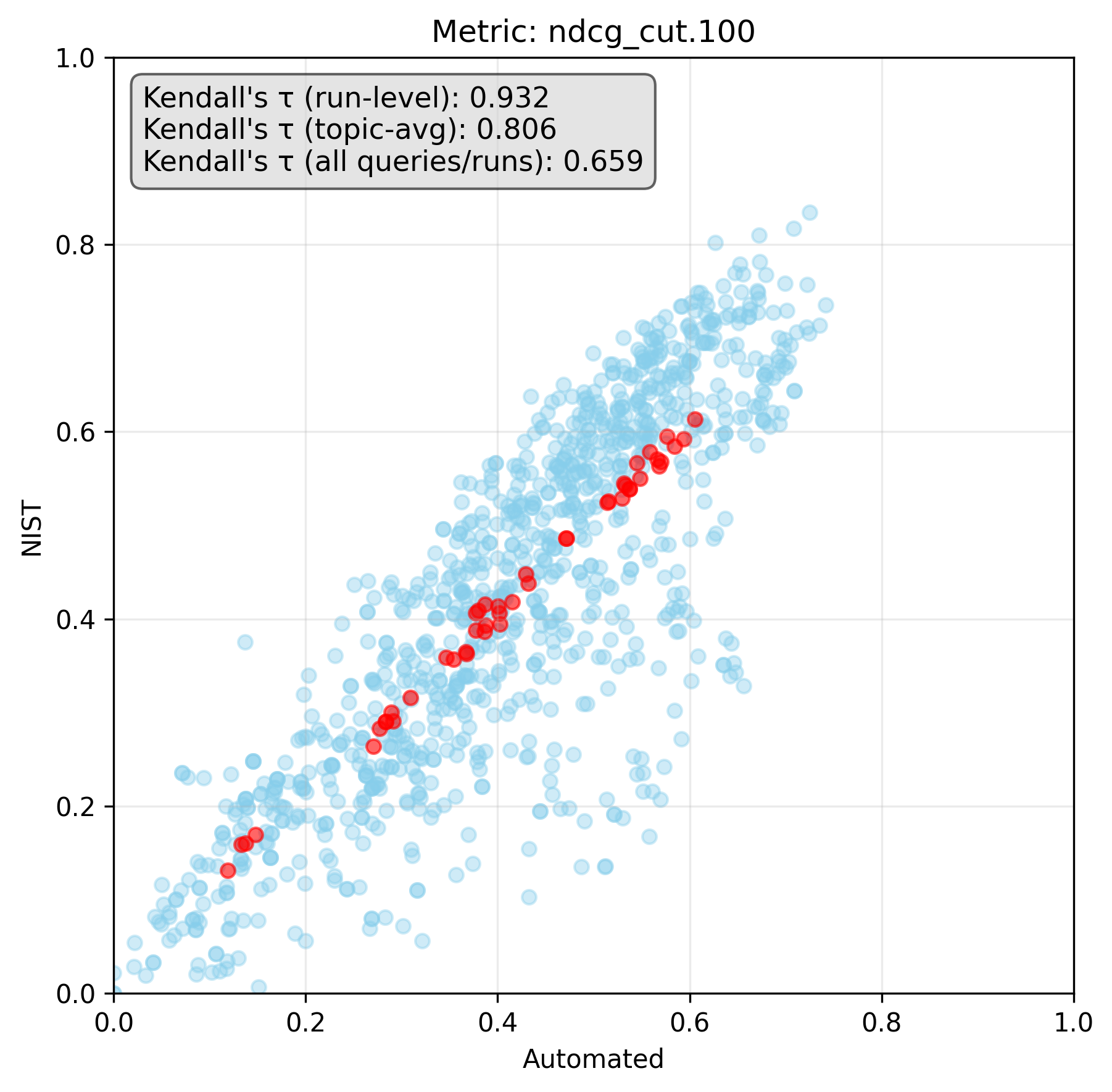}
    \includegraphics[width=0.32\textwidth]{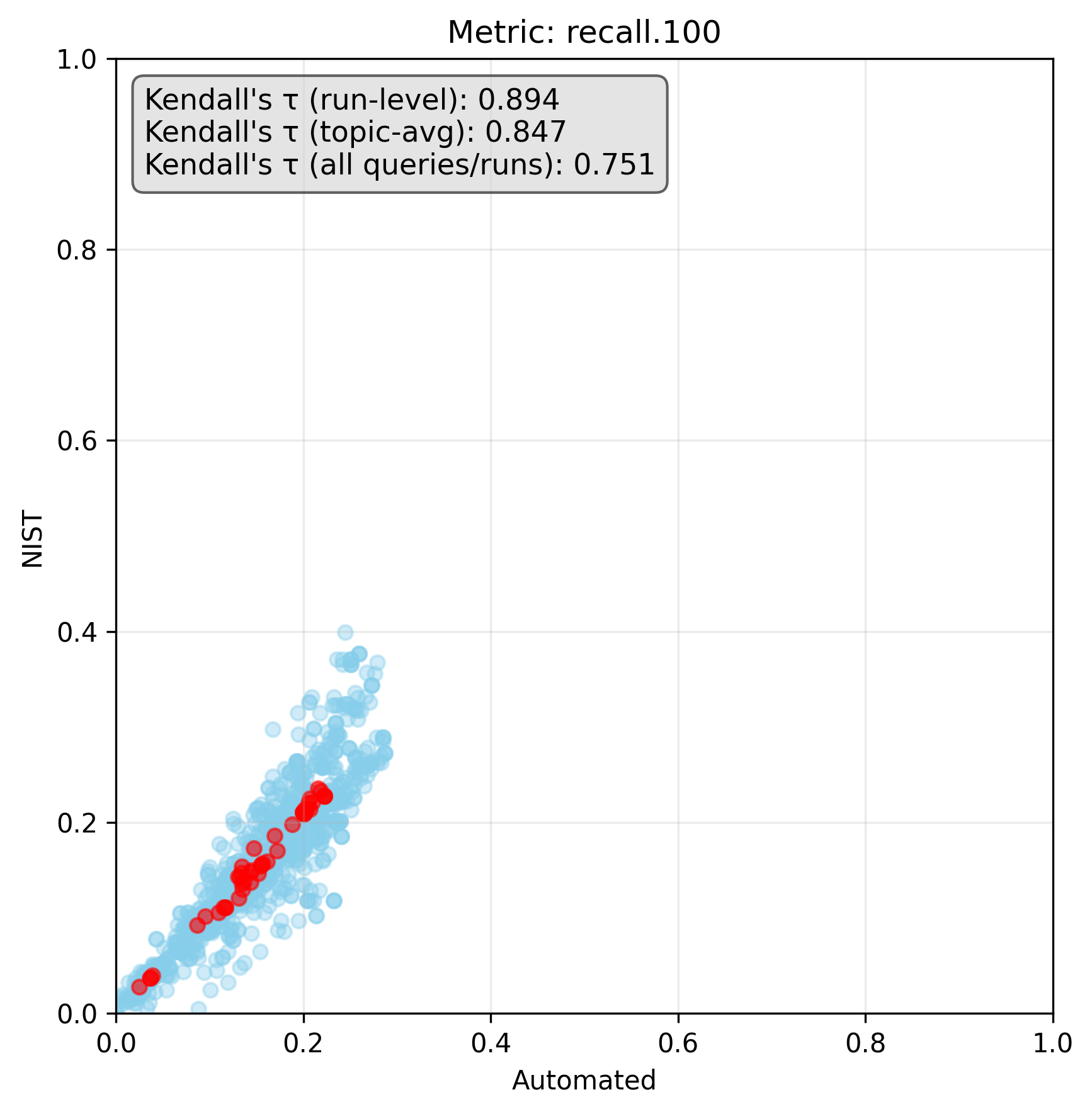} 
    \caption{Rank correlation analysis between automated and manual judgments. Here, Ensemble of models refers to the following combination of models: GPT‑4.1, Gemini 2.5 Pro, Qwen3 Thinking 32B, and GPT‑OSS (20B/120B) with medium level reasoning.}
    \label{fig:retr_analysis}
\end{figure*}


 \begin{figure}[bth]
        \begin{subfigure}{\textwidth}
            \centering
            \includegraphics[width=0.45\textwidth]{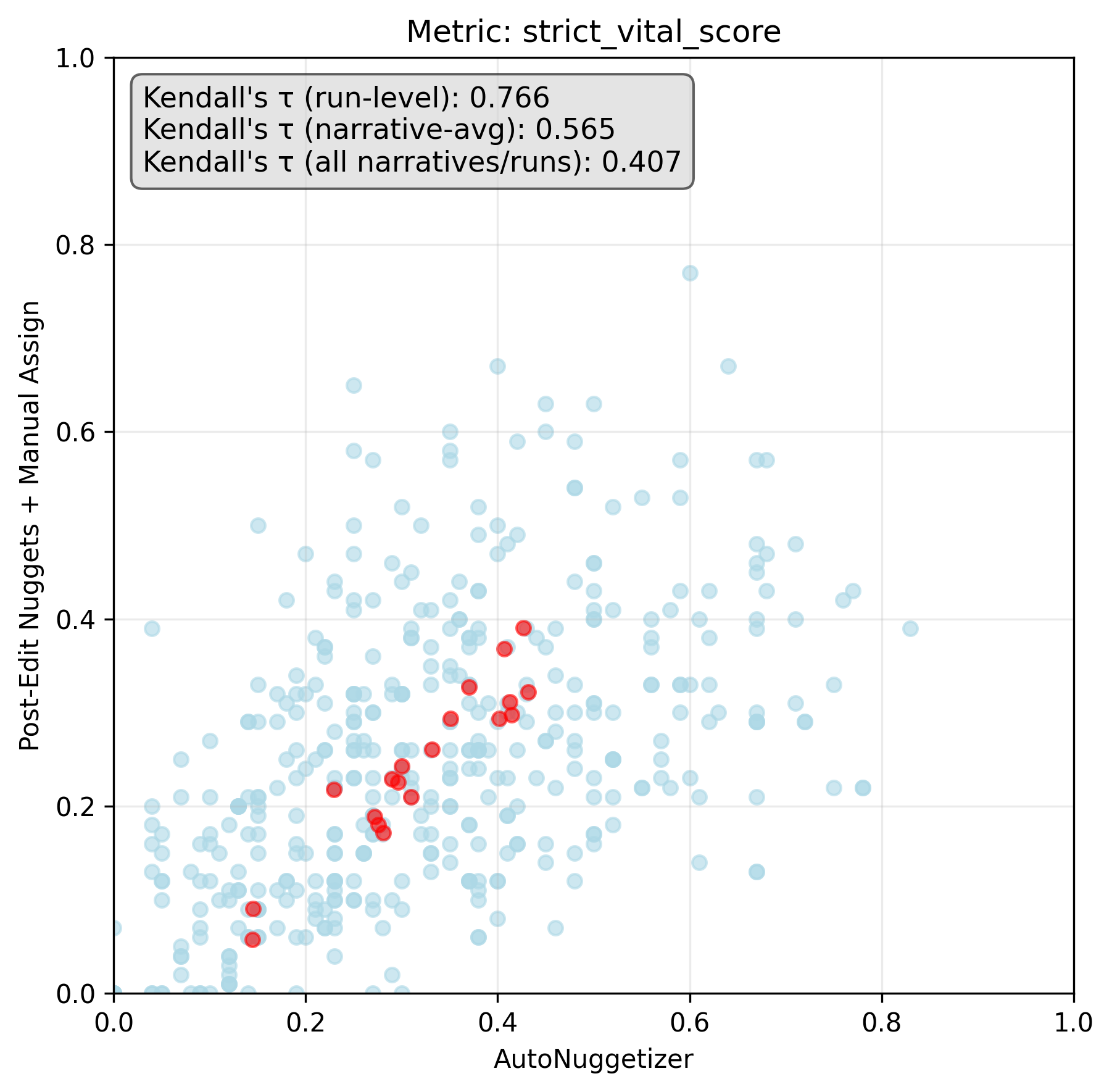}
        \includegraphics[width=0.45\textwidth]{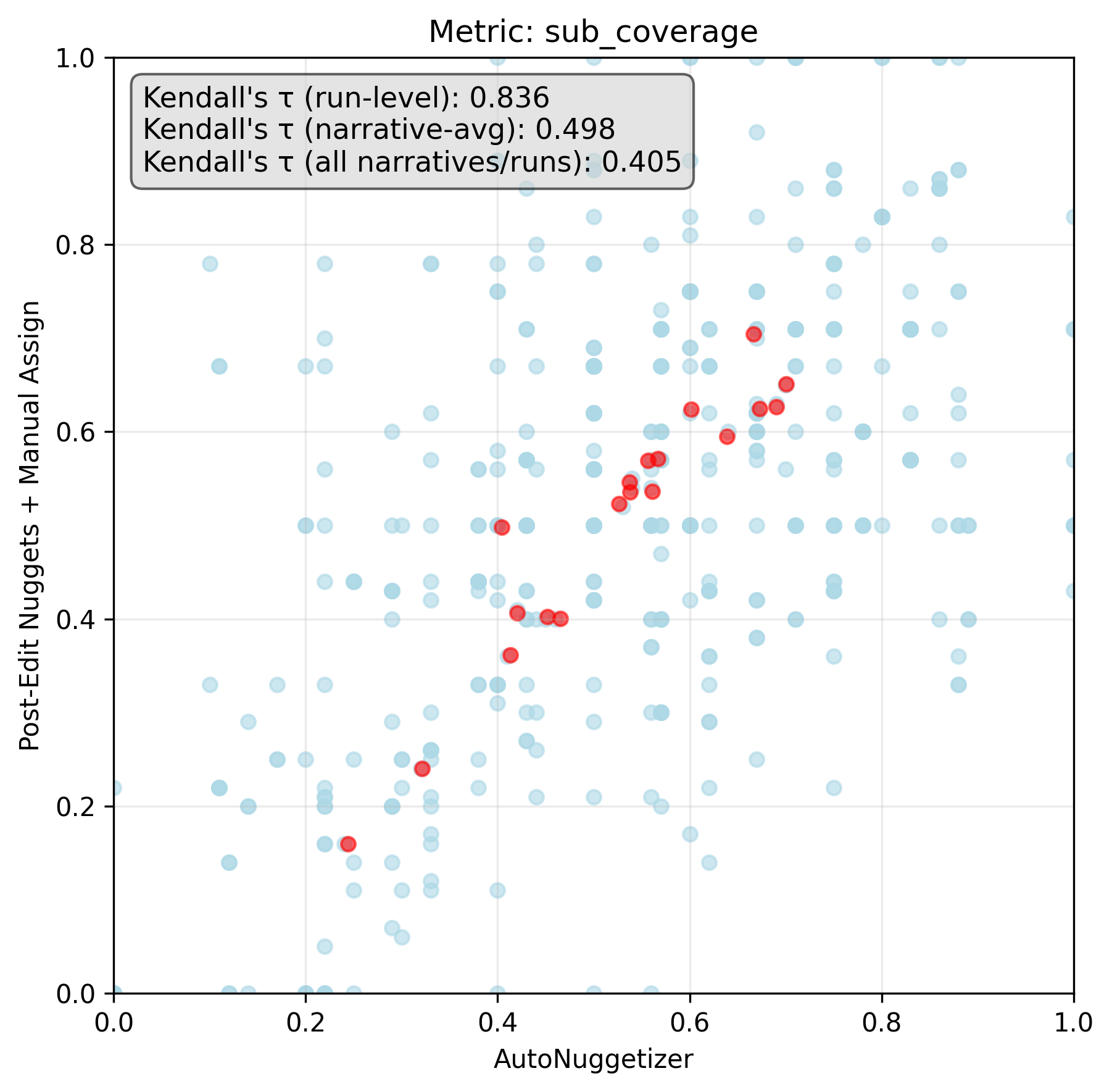}
            \vspace{-0.1cm} 
            \label{fig:row2}
            \caption{Post-Edit Nuggets + ManualAssign vs.\ Fully Automated}
        \end{subfigure} \\
        \begin{subfigure}{\textwidth}
            \centering
            \includegraphics[width=0.45\textwidth]{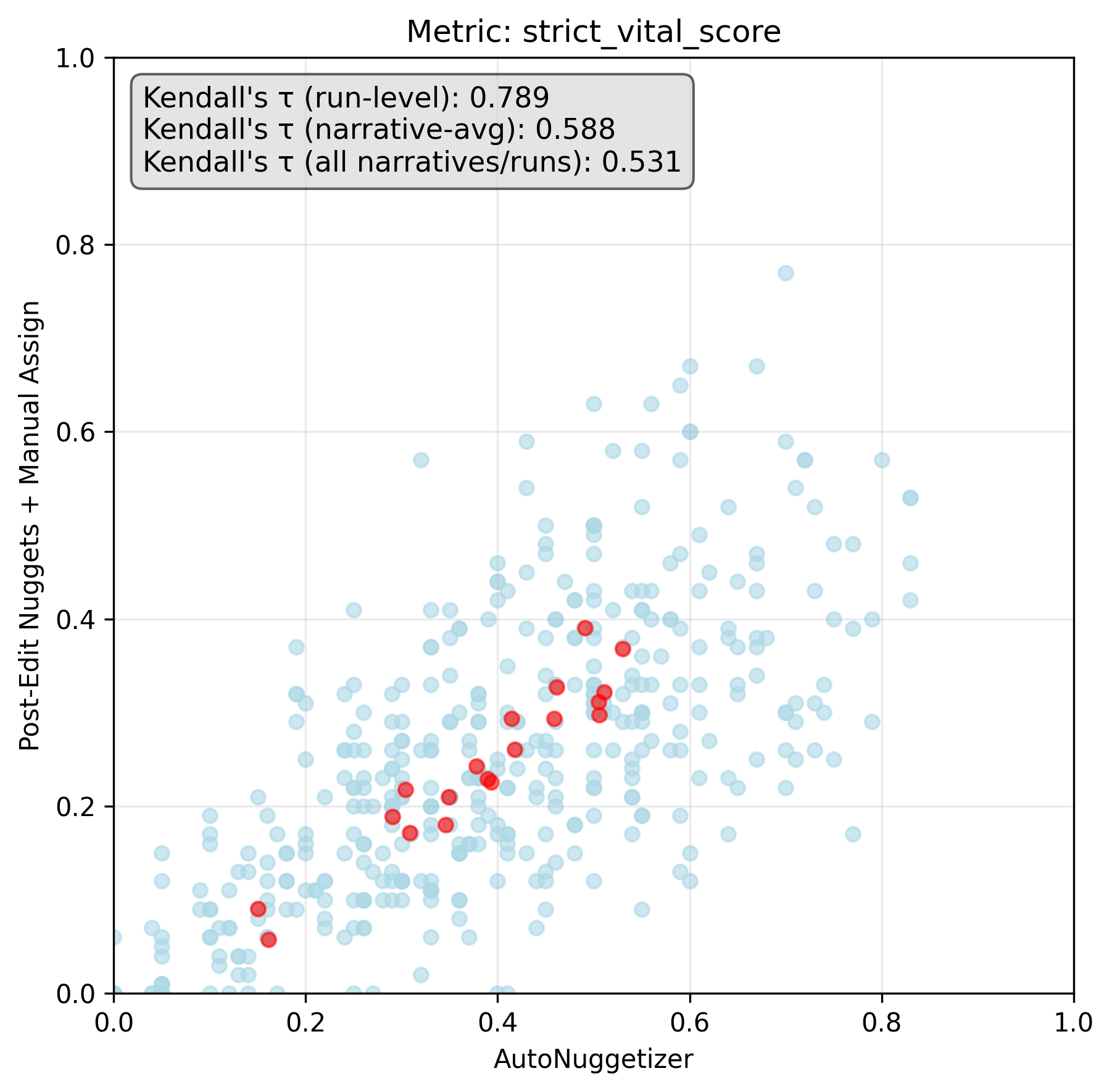}
        \includegraphics[width=0.45\textwidth]{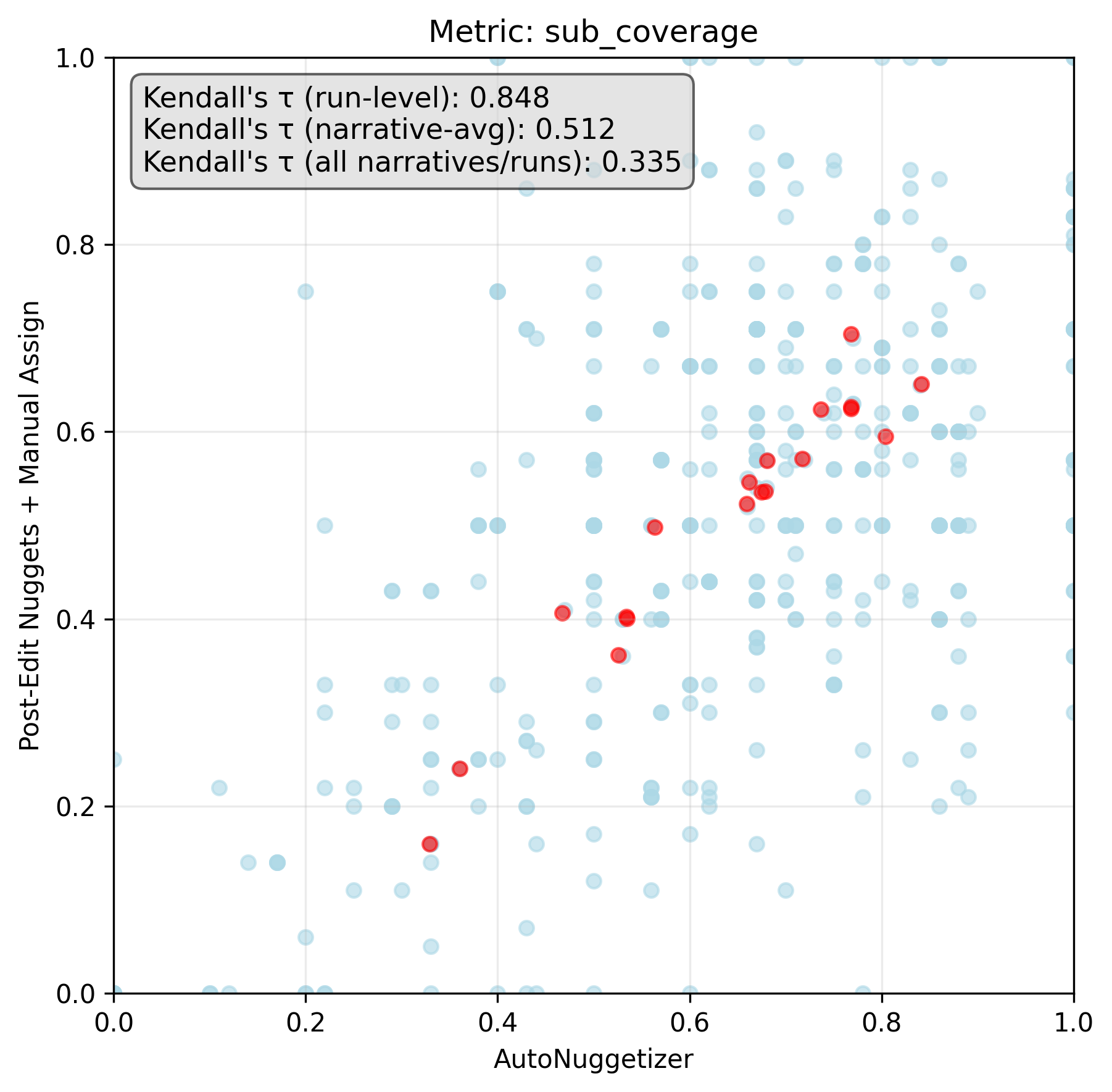}
            \vspace{-0.1cm} 
            \label{fig:row2}
            \caption{Post-Edit Nuggets + ManualAssign vs.\ AutoNuggetizer with NIST qrel}
        \end{subfigure} \\
        \begin{subfigure}{\textwidth}
            \centering
            \includegraphics[width=0.45\textwidth]{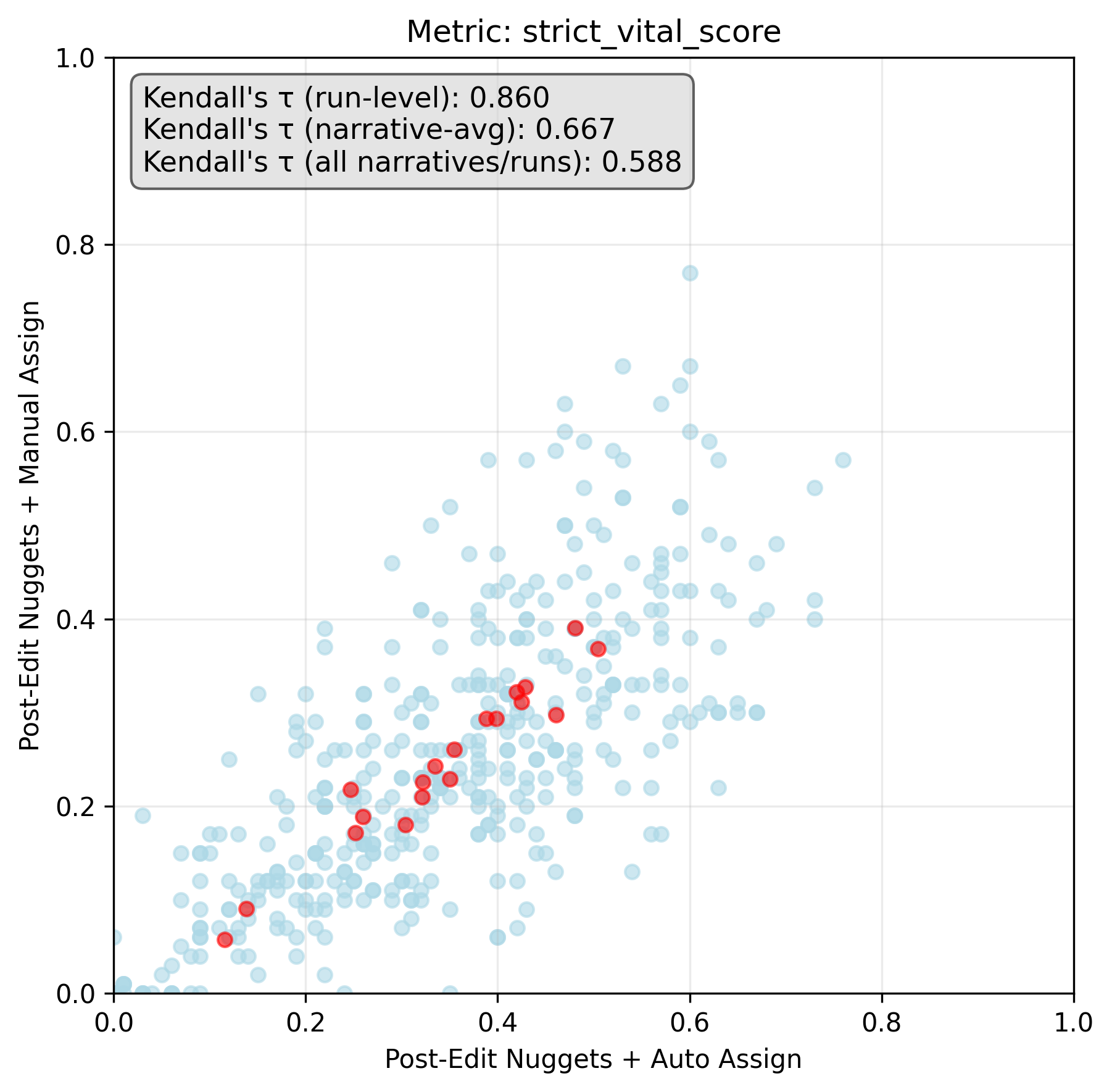}
        \includegraphics[width=0.45\textwidth]{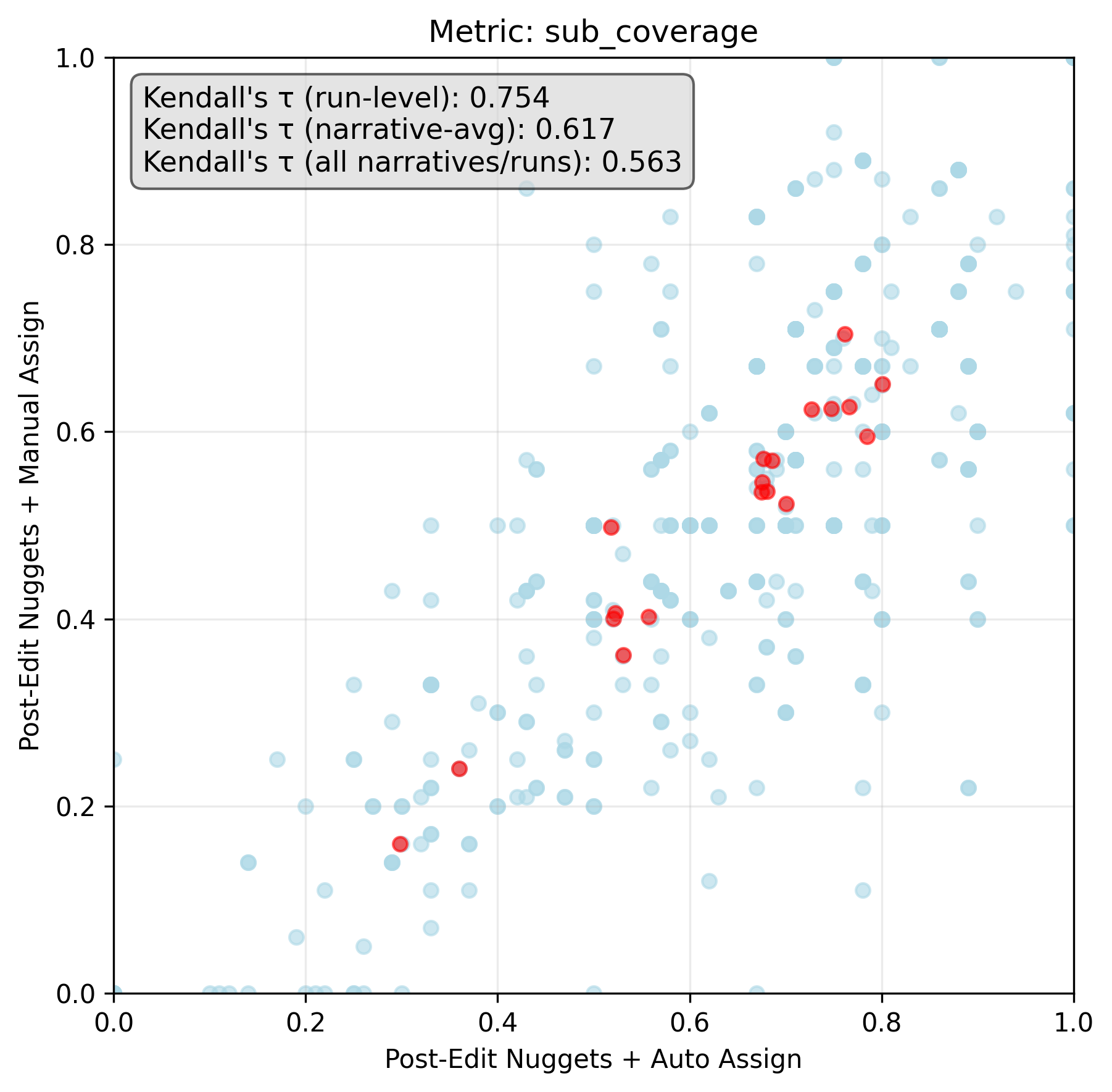}
            \vspace{-0.1cm} 
            \label{fig:row2}
            \caption{Post-Edit Nuggets + ManualAssign vs.\ Post-Edit Nuggets + AutoAssign}
        \end{subfigure}
    \caption{Comparisons between response evaluation scores calculated at varied levels of LLM involvement and manual creation.}
    \label{fig:nugget_analysis1}
\end{figure}

\section{Results}

\subsection{Retrieval Task}
For the TREC 2025 RAG Track, NIST received 46 runs from 12 groups for the Retrieval only task.
Table~\ref{tab:retr_nist} shows scores for Retrieval runs evaluated using NIST manual relevance assessments.
Table~\ref{tab:res_auto} shows scores for Retrieval runs evaluated using automated relevance assessments.
Due to time constraints, the NIST assessors were able to assess documents from only the three highest priority submissions from each group.
In total, these results cover a total of 22 narratives selected by NIST assessors based on their preference.

\begin{table}[t]
\centering
\resizebox{\columnwidth}{!}{%
\begin{tabular}{l l l l ccc}
\hline
Run ID & Group & Run Type & Category-ret & nDCG@30 & nDCG@100 & Recall@100 \\ \hline
4method\_merge & UTokyo & automatic &  Generation-in-the-loop Pipeline & 0.6934 & 0.6134 & 0.2331 \\
lucerank & digsci & automatic &  Generation-in-the-loop Pipeline & 0.6805 & 0.5927 & 0.2274 \\
LAS\_con-que-con-nug & ncsu-las & automatic &  Generation-in-the-loop Pipeline & 0.6692 & 0.4861 & 0.1728 \\
hltcoe-searcher & hltcoe-rerank & automatic &  Multi-Stage Pipeline pointwise & 0.6565 & 0.5786 & 0.2251 \\
qwen\_splade & UTokyo & automatic &  Generation-in-the-loop Pipeline & 0.6517 & 0.5291 & 0.1979 \\
hltcoe-fsrrf & hltcoe-rerank & automatic &  Ensemble/Fusion of First Stages & 0.6463 & 0.5952 & 0.2353 \\
rag25\_test\_rankqwen3 & coordinators & automatic &  Multi-Stage Pipeline pointwise & 0.6462 & 0.5844 & 0.2274 \\
LAS\_con-que-sep-nug & ncsu-las & automatic &  Generation-in-the-loop Pipeline & 0.6447 & 0.4382 & 0.1539 \\
ret-gemma & MITLL & automatic &  Multi-Stage Pipeline pointwise+pair/listwise & 0.6439 & 0.5637 & 0.2139 \\
full-ret & MITLL & automatic &  Multi-Stage Pipeline pointwise+pair/listwise & 0.6250 & 0.5704 & 0.2212 \\
hltcoe-qwen & hltcoe-rerank & automatic &  Multi-Stage Pipeline pointwise & 0.6242 & 0.5668 & 0.2202 \\
ret-no-decomp & MITLL & automatic &  Multi-Stage Pipeline pointwise & 0.6112 & 0.5428 & 0.2105 \\
ret-no-reranker & MITLL & automatic &  Learned Sparse Only & 0.6048 & 0.5501 & 0.2141 \\
splade-v3-arctic-l & coordinators & automatic &  Ensemble/Fusion of First Stages & 0.6009 & 0.5683 & 0.2274 \\
rag25\_test\_splade-v3 & coordinators & automatic &  Learned Sparse Only & 0.5957 & 0.5387 & 0.2103 \\
ret-splade-only & MITLL & automatic &  Learned Sparse Only & 0.5957 & 0.5389 & 0.2105 \\
splade-v3-arctic-m & coordinators & automatic &  Ensemble/Fusion of First Stages & 0.5925 & 0.5448 & 0.2135 \\
grilllab-agent-gpt45 & grilllab & automatic &  Generation-in-the-loop Pipeline & 0.5865 & 0.4093 & 0.1434 \\
grilllab-agentic-gpt4 & grilllab & automatic &  Generation-in-the-loop Pipeline & 0.5762 & 0.4155 & 0.1475 \\
LAS\_con-que & ncsu-las & automatic &  Generation-in-the-loop Pipeline & 0.5545 & 0.4862 & 0.1860 \\
hltcoe-qwen-jina & hltcoe-rerank & automatic &  Multi-Stage Pipeline pointwise & 0.5377 & 0.5260 & 0.2092 \\
hltcoe-jina & hltcoe-rerank & automatic &  Multi-Stage Pipeline pointwise & 0.5360 & 0.5244 & 0.2092 \\
hybrid-rerank & clip2025 & manual &  Learned Dense Only & 0.5346 & 0.4478 & 0.1705 \\
LAS\_sep-que & ncsu-las & automatic &  Generation-in-the-loop Pipeline & 0.5332 & 0.3880 & 0.1427 \\
NITA\_R\_JH\_HY & NITATREC & automatic &  Multi-Stage Pipeline pointwise & 0.5270 & 0.4067 & 0.1497 \\
RRF\_colbert\_minlm & cfdalab & automatic &  Multi-Stage Pipeline pointwise & 0.5262 & 0.2907 & 0.0926 \\
ronly\_combined & WaterlooClarke & automatic &  Generation-in-the-loop Pipeline & 0.5193 & 0.3948 & 0.1463 \\
RRF\_all & cfdalab & automatic &  Multi-Stage Pipeline pointwise & 0.5143 & 0.3631 & 0.1381 \\
rag25\_test\_arctic-m & coordinators & automatic &  Learned Dense Only & 0.5028 & 0.4136 & 0.1559 \\
rag25\_test\_arctic-l & coordinators & automatic &  Learned Dense Only & 0.4991 & 0.4183 & 0.1592 \\
NITA\_R\_DPR & NITATREC & automatic &  Multi-Stage Pipeline pointwise & 0.4864 & 0.3586 & 0.1302 \\
ronly\_nuggetizer & WaterlooClarke & automatic & Traditional Only & 0.4863 & 0.4065 & 0.1481 \\
ronly\_auto\_selected & WaterlooClarke & automatic &  Generation-in-the-loop Pipeline & 0.4599 & 0.3570 & 0.1357 \\
ronly\_garag & WaterlooClarke & automatic &  Generation-in-the-loop Pipeline & 0.4533 & 0.3647 & 0.1372 \\
RRF\_minilm\_bm25 & cfdalab & automatic &  Multi-Stage Pipeline pointwise & 0.4454 & 0.3935 & 0.1545 \\
sentence-transformers-all-MiniLM-L6-v2 & clip2025 & automatic & Learned Dense Only & 0.4029 & 0.3163 & 0.1211 \\
RRF\_colert\_bm25 & cfdalab & automatic &  Multi-Stage Pipeline pointwise & 0.3992 & 0.3863 & 0.1571 \\
bm25\_NITA\_JH & NITATREC & automatic & Traditional Only & 0.3966 & 0.3001 & 0.1112 \\
ronly\_auto\_plan & WaterlooClarke & automatic &  Generation-in-the-loop Pipeline & 0.3760 & 0.2637 & 0.1018 \\
bm25\_rocchio\_top100 & coordinators & automatic & Traditional Only & 0.3634 & 0.2831 & 0.1056 \\
anserini\_bm25\_top100 & coordinators & automatic & Traditional Only & 0.3468 & 0.2900 & 0.1107 \\
Anserini\_bm25\_only & cfdalab & automatic & Traditional Only & 0.3468 & 0.2900 & 0.1107 \\
uema2lab\_rrf\_k10 & tus & automatic &  Ensemble/Fusion of First Stages & 0.3410 & 0.1697 & 0.0396 \\
uema2lab\_rrf & tus & automatic &  Ensemble/Fusion of First Stages & 0.3219 & 0.1601 & 0.0371 \\
uema2lab\_segment & tus & automatic &  Ensemble/Fusion of First Stages & 0.3200 & 0.1592 & 0.0371 \\
uema2lab\_narrative & tus & automatic &  Ensemble/Fusion of First Stages & 0.2646 & 0.1311 & 0.0278 \\
\hline
\end{tabular}}
\vspace{0.3cm}
\caption{Summary of the manual relevance assessment for the retrieval task for 22 narratives.}
\vspace{-0.5cm}
\label{tab:retr_nist}
\end{table}
\begin{table}[t]
\centering
\resizebox{\columnwidth}{!}{%
\begin{tabular}{l l l l ccc}
\hline
Run ID & Group & Run Type & Category-ret & nDCG@30 & nDCG@100 & Recall@100 \\ \hline
4method\_merge & UTokyo & automatic &  Generation-in-the-loop Pipeline & 0.6762 & 0.6052 & 0.2185 \\
lucerank & digsci & automatic &  Generation-in-the-loop Pipeline & 0.6595 & 0.5937 & 0.2226 \\
qwen\_splade & UTokyo & automatic &  Generation-in-the-loop Pipeline & 0.6526 & 0.5296 & 0.1881 \\
LAS\_con-que-con-nug & ncsu-las & automatic &  Generation-in-the-loop Pipeline & 0.6493 & 0.4719 & 0.1471 \\
ret-gemma & MITLL & automatic &  Multi-Stage Pipeline pointwise+pair/listwise & 0.6371 & 0.5679 & 0.2036 \\
LAS\_con-que-sep-nug & ncsu-las & automatic &  Generation-in-the-loop Pipeline & 0.6195 & 0.4321 & 0.1340 \\
rag25\_test\_rankqwen3 & coordinators & automatic &  Multi-Stage Pipeline pointwise & 0.6146 & 0.5845 & 0.2226 \\
hltcoe-searcher & hltcoe-rerank & automatic &  Multi-Stage Pipeline pointwise & 0.6129 & 0.5587 & 0.2069 \\
hltcoe-fsrrf & hltcoe-rerank & automatic &  Ensemble/Fusion of First Stages & 0.6080 & 0.5765 & 0.2157 \\
full-ret & MITLL & automatic &  Multi-Stage Pipeline pointwise+pair/listwise & 0.5981 & 0.5659 & 0.2094 \\
ret-no-reranker & MITLL & automatic &  Learned Sparse Only & 0.5854 & 0.5483 & 0.2016 \\
hltcoe-qwen & hltcoe-rerank & automatic &  Multi-Stage Pipeline pointwise & 0.5846 & 0.5453 & 0.2054 \\
rag25\_test\_splade-v3 & coordinators & automatic &  Learned Sparse Only & 0.5838 & 0.5372 & 0.1992 \\
ret-splade-only & MITLL & automatic &  Learned Sparse Only & 0.5838 & 0.5373 & 0.1993 \\
ret-no-decomp & MITLL & automatic &  Multi-Stage Pipeline pointwise & 0.5754 & 0.5325 & 0.1993 \\
splade-v3-arctic-l & coordinators & automatic &  Ensemble/Fusion of First Stages & 0.5747 & 0.5702 & 0.2226 \\
splade-v3-arctic-m & coordinators & automatic &  Ensemble/Fusion of First Stages & 0.5520 & 0.5315 & 0.2067 \\
grilllab-agent-gpt45 & grilllab & automatic &  Generation-in-the-loop Pipeline & 0.5299 & 0.3797 & 0.1301 \\
LAS\_con-que & ncsu-las & automatic &  Generation-in-the-loop Pipeline & 0.5252 & 0.4710 & 0.1692 \\
grilllab-agentic-gpt4 & grilllab & automatic &  Generation-in-the-loop Pipeline & 0.5249 & 0.3866 & 0.1337 \\
RRF\_colbert\_minlm & cfdalab & automatic &  Multi-Stage Pipeline pointwise & 0.5110 & 0.2912 & 0.0869 \\
RRF\_all & cfdalab & automatic &  Multi-Stage Pipeline pointwise & 0.5054 & 0.3675 & 0.1364 \\
ronly\_combined & WaterlooClarke & automatic &  Generation-in-the-loop Pipeline & 0.5051 & 0.4022 & 0.1516 \\
hltcoe-qwen-jina & hltcoe-rerank & automatic &  Multi-Stage Pipeline pointwise & 0.5042 & 0.5153 & 0.2008 \\
hltcoe-jina & hltcoe-rerank & automatic &  Multi-Stage Pipeline pointwise & 0.5037 & 0.5144 & 0.2006 \\
hybrid-rerank & clip2025 & manual &  Learned Dense Only & 0.4996 & 0.4294 & 0.1717 \\
LAS\_sep-que & ncsu-las & automatic &  Generation-in-the-loop Pipeline & 0.4965 & 0.3769 & 0.1323 \\
rag25\_test\_arctic-l & coordinators & automatic &  Learned Dense Only & 0.4880 & 0.4150 & 0.1614 \\
rag25\_test\_arctic-m & coordinators & automatic &  Learned Dense Only & 0.4830 & 0.4003 & 0.1552 \\
NITA\_R\_JH\_HY & NITATREC & automatic &  Multi-Stage Pipeline pointwise & 0.4778 & 0.3774 & 0.1444 \\
ronly\_nuggetizer & WaterlooClarke & automatic & Traditional Only & 0.4701 & 0.4015 & 0.1436 \\
NITA\_R\_DPR & NITATREC & automatic &  Multi-Stage Pipeline pointwise & 0.4522 & 0.3461 & 0.1352 \\
ronly\_garag & WaterlooClarke & automatic &  Generation-in-the-loop Pipeline & 0.4415 & 0.3668 & 0.1437 \\
ronly\_auto\_selected & WaterlooClarke & automatic &  Generation-in-the-loop Pipeline & 0.4364 & 0.3543 & 0.1344 \\
RRF\_minilm\_bm25 & cfdalab & automatic &  Multi-Stage Pipeline pointwise & 0.4259 & 0.3878 & 0.1543 \\
RRF\_colert\_bm25 & cfdalab & automatic &  Multi-Stage Pipeline pointwise & 0.3893 & 0.3859 & 0.1568 \\
sentence-transformers-all-MiniLM-L6-v2 & clip2025 & automatic & Learned Dense Only & 0.3848 & 0.3089 & 0.1311 \\
ronly\_auto\_plan & WaterlooClarke & automatic &  Generation-in-the-loop Pipeline & 0.3749 & 0.2707 & 0.0955 \\
bm25\_NITA\_JH & NITATREC & automatic & Traditional Only & 0.3665 & 0.2890 & 0.1150 \\
bm25\_rocchio\_top100 & coordinators & automatic & Traditional Only & 0.3306 & 0.2767 & 0.1092 \\
anserini\_bm25\_top100 & coordinators & automatic & Traditional Only & 0.3250 & 0.2835 & 0.1167 \\
Anserini\_bm25\_only & cfdalab & automatic & Traditional Only & 0.3250 & 0.2835 & 0.1167 \\
uema2lab\_rrf\_k10 & tus & automatic &  Ensemble/Fusion of First Stages & 0.2916 & 0.1475 & 0.0390 \\
uema2lab\_rrf & tus & automatic &  Ensemble/Fusion of First Stages & 0.2722 & 0.1376 & 0.0368 \\
uema2lab\_segment & tus & automatic &  Ensemble/Fusion of First Stages & 0.2635 & 0.1331 & 0.0362 \\
uema2lab\_narrative & tus & automatic &  Ensemble/Fusion of First Stages & 0.2356 & 0.1188 & 0.0244 \\
\hline
\end{tabular}}
\vspace{0.3cm}
\caption{Summary of the automatic relevance assessment for the retrieval task for 22 narratives.}
\label{tab:retr_auto}
\end{table}

At a high level, our analyses examine the correlation between system rankings derived from evaluation scores computed using automated artifacts and those induced through manual assessment.
In accordance with standard practice in Information Retrieval meta‑evaluation, rank agreement is quantified using Kendall’s $\tau$ correlation coefficient.
The experimental design, which follows~\citet{trec24_ret} can be explained as follows:
\begin{itemize}[leftmargin=*]
\item \textbf{Run-level correlations on shared narratives (run-level).}  
For each submitted run, system effectiveness is evaluated using two sets of artifacts: manual assessments and automatically generated assessments for the same narrative set. 
Kendall’s $\tau$ is then computed between the resulting effectiveness scores.

\item \textbf{Mean of per-narrative correlations (narrative-avg).}  
As above, each run is evaluated using both the manual assessments and the corresponding automatically generated assessments over the same manually judged narratives.  
In this setting, however, Kendall’s $\tau$ is calculated separately for each narrative, and the final correlation value is obtained by averaging these per-narrative correlations.

\item \textbf{Correlation over all narrative--run pairs (all narratives~/~runs).}  
Using the same automatically generated assessments as in the preceding settings, each narrative--run pair is treated as an independent observation.  
Kendall’s $\tau$ is computed over the complete set of such observations.
\end{itemize}

The Figure~\ref{fig:retr_analysis} presents a comparative analysis of manual evaluation scores and the outputs of the ensemble across the nDCG@30, nDCG@100, and recall@100 metrics. 
In all cases, the automated assessments demonstrate a strong ability to approximate manual judgments at the run level, as evidenced by the high Kendall’s $\tau$ coefficients and the close alignment of the aggregated red points with the identity line.
This indicates that, when scores are aggregated across entire runs, the automated system captures the overall ranking behavior produced by human assessors with considerable fidelity. 
At finer levels of granularity, such as narrative-avg and all narratives~/~runs comparisons, the scatter becomes more pronounced, suggesting greater variability in how individual narratives are evaluated and highlighting inherent noise in localized relevance patterns. 
Notably, the degree of correspondence improves at larger cutoff values, particularly for nDCG@100, where the distribution tightens, and the rank correlations increase. 
This trend suggests that the ensemble yields more stable approximations when the metric incorporates a broader portion of the ranked list. 
Overall, the results indicate that the automated method provides a reliable run‑level surrogate for manual evaluation, while still exhibiting expected deviations at the narrative level.

\begin{table}[t]
\centering
\scalebox{0.65}{
\begin{tabular}{l l l cc}
\hline
Run ID & Group & Task & strict vital score & sub-narrative coverage \\ \hline
Kun-Third & RMIT-IR & RAG & 0.3900 & 0.7000 \\
LAS-agentic-RAG-agent & ncsu-las & RAG & 0.3700 & 0.6500 \\
uema2lab\_B4 & tus & RAG & 0.3300 & 0.5500 \\
combined & WaterlooClarke & RAG & 0.3200 & 0.6300 \\
auto\_selected & WaterlooClarke & RAG & 0.3100 & 0.6300 \\
LAS-agentic-RAG-selector & ncsu-las & RAG & 0.3000 & 0.6000 \\
IDACCS-nugg-gpt-4-1 & IDACCS & AG & 0.2900 & 0.5700 \\
auto\_plan & WaterlooClarke & RAG & 0.2900 & 0.6200 \\
uema2lab\_base & tus & RAG & 0.2700 & 0.5300 \\
KG-AG-1 & clip2025 & RAG & 0.2600 & 0.5700 \\
IDACCSabstrct-gpt4-1 & IDACCS & AG & 0.2400 & 0.5400 \\
sub\_query\_entities & clip2025 & RAG & 0.2300 & 0.5400 \\
ori\_query\_entities & clip2025 & RAG & 0.2300 & 0.5200 \\
genSubQ\_merge & uogTr & RAG & 0.2200 & 0.5000 \\
wingii-v3-gpt & WING-II & AG & 0.2100 & 0.4000 \\
IDACCS-hybrid-gpt4-1 & IDACCS & AG & 0.1900 & 0.4000 \\
wingii-3-rl-refined & WING-II & AG & 0.1800 & 0.3600 \\
uema2lab\_rag\_fewdoc & tus & RAG & 0.1700 & 0.4100 \\
e5\_monot5\_searchR1 & uogTr & RAG & 0.0900 & 0.2400 \\
no-llm-refined & WING-II & AG & 0.0600 & 0.1600 \\
\hline
\end{tabular}}
\vspace{0.3cm}
\caption{Summary of the RAG and AG task scores of runs with priority $\geq$ 3 with nugget post-editing with manual assignment for 22 narratives.}
\label{tab:res_manual_assign}
\vspace{-0.5cm}
\end{table}

\begin{table}[t]
\centering
\scalebox{0.65}{
\begin{tabular}{l l l cc}
\hline
Run ID & Group & Task & strict vital score & sub-narrative coverage \\ \hline
LAS-agentic-RAG-agent & ncsu-las & RAG & 0.5000 & 0.8000 \\
no-decomp & MITLL & RAG & 0.5000 & 0.7700 \\
r\_4method\_ag\_gpt41 & UTokyo & RAG & 0.4900 & 0.8100 \\
gpt41 & UTokyo & AG & 0.4800 & 0.8100 \\
Kun-Third & RMIT-IR & RAG & 0.4800 & 0.7600 \\
no-reranker & MITLL & RAG & 0.4800 & 0.7700 \\
rag25\_qwen3\_50\_ag & coordinators & AG & 0.4700 & 0.8000 \\
full & MITLL & RAG & 0.4700 & 0.7900 \\
r\_2method\_ag\_gpt41 & UTokyo & RAG & 0.4700 & 0.8200 \\
rag25\_test\_qwen3\_50 & coordinators & RAG & 0.4700 & 0.8100 \\
hltcoe-gpt5.searcher & hltcoe-rerank & RAG & 0.4700 & 0.8100 \\
single-agent-trim & ncsu-las & RAG & 0.4700 & 0.7400 \\
LAS-agentic-RAG-selector & ncsu-las & RAG & 0.4600 & 0.7800 \\
rag25\_qwen3\_20\_ag & coordinators & AG & 0.4500 & 0.7800 \\
rag25\_test\_qwen3\_20 & coordinators & RAG & 0.4500 & 0.7900 \\
no-decomp-reranker & MITLL & RAG & 0.4400 & 0.7400 \\
selector-agent-trim & ncsu-las & RAG & 0.4300 & 0.7200 \\
uema2lab\_B4 & tus & RAG & 0.4300 & 0.6800 \\
rag\_v4 & cfdalab & RAG & 0.4300 & 0.7300 \\
auto\_selected & WaterlooClarke & RAG & 0.4200 & 0.7700 \\
combined & WaterlooClarke & RAG & 0.4200 & 0.7500 \\
IDACCS-nugg-gpt-4-1 & IDACCS & AG & 0.4000 & 0.6900 \\
Rerank-Top50\_v3 & clip2025 & AG & 0.4000 & 0.7100 \\
garag & WaterlooClarke & RAG & 0.4000 & 0.7000 \\
auto\_plan & WaterlooClarke & RAG & 0.3900 & 0.7300 \\
nugget-generation & GenAIus & AG & 0.3800 & 0.6600 \\
ag-v2-gpt & cfdalab & AG & 0.3800 & 0.7000 \\
cluster-generation & GenAIus & AG & 0.3800 & 0.6400 \\
ag-v1-gpt & cfdalab & AG & 0.3800 & 0.7200 \\
rag-v1-gpt & cfdalab & RAG & 0.3800 & 0.7100 \\
rag-v2-gpt & cfdalab & RAG & 0.3700 & 0.7200 \\
Rerank-Top50\_v2 & clip2025 & AG & 0.3600 & 0.6700 \\
KG-AG-1 & clip2025 & RAG & 0.3600 & 0.6800 \\
grilllab-gpt45-gen & grilllab & RAG & 0.3600 & 0.7000 \\
uema2lab\_rag\_org & tus & RAG & 0.3600 & 0.7000 \\
ori\_query\_entities & clip2025 & RAG & 0.3500 & 0.7000 \\
nuggetizer & WaterlooClarke & RAG & 0.3400 & 0.7000 \\
grilllab-agentic-gpt4-generation & grilllab & RAG & 0.3400 & 0.6900 \\
IDACCSabstrct-gpt4-1 & IDACCS & AG & 0.3300 & 0.6800 \\
ag-v2-llama & cfdalab & AG & 0.3300 & 0.6500 \\
wingii-v3-gpt & WING-II & AG & 0.3200 & 0.5600 \\
extractive\_rag & hltcoe-multiagt & RAG & 0.3200 & 0.6200 \\
sub\_query\_entities & clip2025 & RAG & 0.3200 & 0.6700 \\
uema2lab\_base & tus & RAG & 0.3100 & 0.5900 \\
wingii-3-rl-refined & WING-II & AG & 0.3000 & 0.5300 \\
gptr.nt\_q4d4 & hltcoe-multiagt & RAG & 0.3000 & 0.6400 \\
gptr\_e2\_q4d4 & hltcoe-multiagt & RAG & 0.2900 & 0.6300 \\
hltcoe-lg.qwen & hltcoe-rerank & RAG & 0.2900 & 0.6000 \\
hltcoe-lg.jina & hltcoe-rerank & RAG & 0.2900 & 0.5800 \\
lg\_nt\_q4d12l3\_c & hltcoe-multiagt & RAG & 0.2900 & 0.6100 \\
gptr\_nt\_q3d3 & hltcoe-multiagt & RAG & 0.2900 & 0.6000 \\
IDACCS-nugg-gpt-4o & IDACCS & AG & 0.2800 & 0.6000 \\
lg\_nt\_q4d12l3 & hltcoe-multiagt & RAG & 0.2800 & 0.6000 \\
hltcoe-lg.searcher & hltcoe-rerank & RAG & 0.2800 & 0.6200 \\
cru-ansR-bareconf & HLTCOE & AG & 0.2700 & 0.5300 \\
swarm\_c & hltcoe-multiagt & RAG & 0.2700 & 0.6100 \\
hltcoe-lg.fsrrf & hltcoe-rerank & RAG & 0.2700 & 0.5800 \\
swarm & hltcoe-multiagt & RAG & 0.2700 & 0.5900 \\
rag-v2-llama & cfdalab & RAG & 0.2700 & 0.5500 \\
IDACCS-hybrid-gpt4-1 & IDACCS & AG & 0.2600 & 0.5200 \\
cru-ansR-conf & HLTCOE & AG & 0.2600 & 0.5100 \\
cru-ansR & HLTCOE & AG & 0.2600 & 0.5400 \\
IDACCS-hybrid-gpt4o & IDACCS & AG & 0.2500 & 0.5100 \\
uema2lab\_rag\_fewdoc & tus & RAG & 0.2500 & 0.5200 \\
genSubQ\_merge & uogTr & RAG & 0.2500 & 0.5200 \\
cru-ablR & HLTCOE & AG & 0.2300 & 0.4800 \\
gptr\_e2\_q3d3 & hltcoe-multiagt & RAG & 0.2300 & 0.5700 \\
cru-ablR-conf & HLTCOE & AG & 0.2000 & 0.4900 \\
NITA\_AG\_JH & NITATREC & AG & 0.1700 & 0.3800 \\
ag-run-1-JH & NITATREC & RAG & 0.1600 & 0.3800 \\
no-llm & WING-II & AG & 0.1500 & 0.3700 \\
bm25-rz7b-2025a & ii\_research & RAG & 0.1400 & 0.2900 \\
e5\_monot5\_searchR1 & uogTr & RAG & 0.1400 & 0.3600 \\
no-llm-refined & WING-II & AG & 0.1200 & 0.3000 \\
strd\_roll\_segment & IRIT-ISIR-EV & RAG & 0.1200 & 0.2300 \\
standard\_roll & IRIT-ISIR-EV & RAG & 0.1000 & 0.2700 \\
\hline
\end{tabular}}
\vspace{0.3cm}
\caption{Summary of the RAG and AG task scores with nugget post-editing + automatic assignment for 22 narratives.}
\label{tab:res_pe_auto}
\end{table}
\begin{table}[t]
\centering
\scalebox{0.65}{
\begin{tabular}{l l l cc}
\hline
Run ID & Group & Task & strict vital score & sub-narrative coverage \\ \hline
r\_2method\_ag\_gpt41 & UTokyo & RAG & 0.5500 & 0.8400 \\
rag25\_qwen3\_50\_ag & coordinators & AG & 0.5400 & 0.8300 \\
gpt41 & UTokyo & AG & 0.5300 & 0.8400 \\
LAS-agentic-RAG-agent & ncsu-las & RAG & 0.5300 & 0.8400 \\
full & MITLL & RAG & 0.5300 & 0.7600 \\
rag25\_test\_qwen3\_50 & coordinators & RAG & 0.5300 & 0.8100 \\
hltcoe-gpt5.searcher & hltcoe-rerank & RAG & 0.5300 & 0.8000 \\
no-reranker & MITLL & RAG & 0.5200 & 0.7400 \\
no-decomp & MITLL & RAG & 0.5200 & 0.7600 \\
r\_4method\_ag\_gpt41 & UTokyo & RAG & 0.5200 & 0.7900 \\
rag25\_qwen3\_20\_ag & coordinators & AG & 0.5100 & 0.7900 \\
LAS-agentic-RAG-selector & ncsu-las & RAG & 0.5100 & 0.8000 \\
no-decomp-reranker & MITLL & RAG & 0.5100 & 0.7300 \\
auto\_selected & WaterlooClarke & RAG & 0.5100 & 0.7700 \\
combined & WaterlooClarke & RAG & 0.5100 & 0.7700 \\
rag25\_test\_qwen3\_20 & coordinators & RAG & 0.5000 & 0.7700 \\
Kun-Third & RMIT-IR & RAG & 0.4900 & 0.7700 \\
rag\_v4 & cfdalab & RAG & 0.4800 & 0.7800 \\
Rerank-Top50\_v3 & clip2025 & AG & 0.4600 & 0.7400 \\
uema2lab\_B4 & tus & RAG & 0.4600 & 0.6600 \\
auto\_plan & WaterlooClarke & RAG & 0.4600 & 0.7400 \\
single-agent-trim & ncsu-las & RAG & 0.4600 & 0.7600 \\
selector-agent-trim & ncsu-las & RAG & 0.4500 & 0.7200 \\
Rerank-Top50\_v2 & clip2025 & AG & 0.4400 & 0.7400 \\
ag-v1-gpt & cfdalab & AG & 0.4300 & 0.7700 \\
garag & WaterlooClarke & RAG & 0.4300 & 0.6600 \\
ag-v2-gpt & cfdalab & AG & 0.4200 & 0.7300 \\
KG-AG-1 & clip2025 & RAG & 0.4200 & 0.7200 \\
rag-v2-gpt & cfdalab & RAG & 0.4200 & 0.7500 \\
uema2lab\_rag\_org & tus & RAG & 0.4200 & 0.6600 \\
nugget-generation & GenAIus & AG & 0.4100 & 0.6300 \\
IDACCS-nugg-gpt-4-1 & IDACCS & AG & 0.4100 & 0.6800 \\
grilllab-gpt45-gen & grilllab & RAG & 0.4100 & 0.6800 \\
cluster-generation & GenAIus & AG & 0.4000 & 0.6200 \\
rag-v1-gpt & cfdalab & RAG & 0.4000 & 0.7300 \\
grilllab-agentic-gpt4-generation & grilllab & RAG & 0.4000 & 0.7200 \\
ori\_query\_entities & clip2025 & RAG & 0.3900 & 0.6600 \\
nuggetizer & WaterlooClarke & RAG & 0.3900 & 0.7000 \\
extractive\_rag & hltcoe-multiagt & RAG & 0.3900 & 0.6900 \\
sub\_query\_entities & clip2025 & RAG & 0.3900 & 0.6700 \\
IDACCSabstrct-gpt4-1 & IDACCS & AG & 0.3800 & 0.6800 \\
uema2lab\_base & tus & RAG & 0.3800 & 0.6000 \\
ag-v2-llama & cfdalab & AG & 0.3600 & 0.6600 \\
wingii-v3-gpt & WING-II & AG & 0.3500 & 0.5300 \\
wingii-3-rl-refined & WING-II & AG & 0.3500 & 0.5300 \\
gptr\_nt\_q3d3 & hltcoe-multiagt & RAG & 0.3500 & 0.5800 \\
IDACCS-nugg-gpt-4o & IDACCS & AG & 0.3400 & 0.6000 \\
hltcoe-lg.searcher & hltcoe-rerank & RAG & 0.3400 & 0.6500 \\
swarm & hltcoe-multiagt & RAG & 0.3400 & 0.5300 \\
lg\_nt\_q4d12l3 & hltcoe-multiagt & RAG & 0.3300 & 0.5600 \\
swarm\_c & hltcoe-multiagt & RAG & 0.3300 & 0.5600 \\
gptr\_e2\_q4d4 & hltcoe-multiagt & RAG & 0.3300 & 0.6200 \\
hltcoe-lg.qwen & hltcoe-rerank & RAG & 0.3300 & 0.5700 \\
gptr.nt\_q4d4 & hltcoe-multiagt & RAG & 0.3300 & 0.5900 \\
cru-ansR & HLTCOE & AG & 0.3200 & 0.5300 \\
hltcoe-lg.jina & hltcoe-rerank & RAG & 0.3200 & 0.5700 \\
lg\_nt\_q4d12l3\_c & hltcoe-multiagt & RAG & 0.3200 & 0.5700 \\
cru-ansR-conf & HLTCOE & AG & 0.3100 & 0.5200 \\
cru-ansR-bareconf & HLTCOE & AG & 0.3100 & 0.5100 \\
uema2lab\_rag\_fewdoc & tus & RAG & 0.3100 & 0.4700 \\
hltcoe-lg.fsrrf & hltcoe-rerank & RAG & 0.3100 & 0.5900 \\
IDACCS-hybrid-gpt4o & IDACCS & AG & 0.3000 & 0.5300 \\
genSubQ\_merge & uogTr & RAG & 0.3000 & 0.5600 \\
rag-v2-llama & cfdalab & RAG & 0.3000 & 0.5500 \\
IDACCS-hybrid-gpt4-1 & IDACCS & AG & 0.2900 & 0.5300 \\
gptr\_e2\_q3d3 & hltcoe-multiagt & RAG & 0.2900 & 0.5700 \\
cru-ablR & HLTCOE & AG & 0.2600 & 0.4400 \\
cru-ablR-conf & HLTCOE & AG & 0.2500 & 0.4600 \\
ag-run-1-JH & NITATREC & RAG & 0.1900 & 0.3700 \\
NITA\_AG\_JH & NITATREC & AG & 0.1800 & 0.3500 \\
no-llm & WING-II & AG & 0.1700 & 0.3700 \\
no-llm-refined & WING-II & AG & 0.1600 & 0.3300 \\
bm25-rz7b-2025a & ii\_research & RAG & 0.1500 & 0.2400 \\
e5\_monot5\_searchR1 & uogTr & RAG & 0.1500 & 0.3600 \\
strd\_roll\_segment & IRIT-ISIR-EV & RAG & 0.1300 & 0.2400 \\
standard\_roll & IRIT-ISIR-EV & RAG & 0.1300 & 0.2600 \\
\hline
\end{tabular}}
\vspace{0.3cm}
\caption{Summary of the RAG and AG task scores with AutoNuggetizer for 22 narratives using NIST relevance judgments.}
\label{tab:res_auto}
\end{table}

\begin{table}[t]
\centering
\scalebox{0.65}{
\begin{tabular}{l l l cc}
\hline
RUN ID & Group & Task & strict vital score & sub-narrative coverage \\ \hline
Kun-Third & RMIT-IR & RAG & 0.4300 & 0.6700 \\
full & MITLL & RAG & 0.4300 & 0.6100 \\
no-reranker & MITLL & RAG & 0.4300 & 0.6200 \\
combined & WaterlooClarke & RAG & 0.4300 & 0.6700 \\
no-decomp & MITLL & RAG & 0.4200 & 0.6300 \\
r\_4method\_ag\_gpt41 & UTokyo & RAG & 0.4200 & 0.6600 \\
LAS-agentic-RAG-agent & ncsu-las & RAG & 0.4100 & 0.7000 \\
LAS-agentic-RAG-selector & ncsu-las & RAG & 0.4100 & 0.6400 \\
no-decomp-reranker & MITLL & RAG & 0.4100 & 0.6100 \\
r\_2method\_ag\_gpt41 & UTokyo & RAG & 0.4100 & 0.6300 \\
auto\_selected & WaterlooClarke & RAG & 0.4100 & 0.6900 \\
gpt41 & UTokyo & AG & 0.4000 & 0.6300 \\
rag25\_test\_qwen3\_50 & coordinators & RAG & 0.4000 & 0.6700 \\
auto\_plan & WaterlooClarke & RAG & 0.4000 & 0.6000 \\
hltcoe-gpt5.searcher & hltcoe-rerank & RAG & 0.4000 & 0.6700 \\
rag25\_qwen3\_50\_ag & coordinators & AG & 0.3900 & 0.6600 \\
selector-agent-trim & ncsu-las & RAG & 0.3900 & 0.5800 \\
garag & WaterlooClarke & RAG & 0.3900 & 0.6300 \\
rag25\_qwen3\_20\_ag & coordinators & AG & 0.3800 & 0.5900 \\
rag25\_test\_qwen3\_20 & coordinators & RAG & 0.3800 & 0.6000 \\
rag\_v4 & cfdalab & RAG & 0.3800 & 0.6400 \\
uema2lab\_B4 & tus & RAG & 0.3700 & 0.5400 \\
single-agent-trim & ncsu-las & RAG & 0.3700 & 0.6400 \\
nugget-generation & GenAIus & AG & 0.3600 & 0.5300 \\
uema2lab\_base & tus & RAG & 0.3600 & 0.5300 \\
IDACCS-nugg-gpt-4-1 & IDACCS & AG & 0.3500 & 0.5600 \\
uema2lab\_rag\_org & tus & RAG & 0.3500 & 0.5500 \\
Rerank-Top50\_v3 & clip2025 & AG & 0.3400 & 0.5700 \\
cluster-generation & GenAIus & AG & 0.3400 & 0.5400 \\
Rerank-Top50\_v2 & clip2025 & AG & 0.3400 & 0.6100 \\
KG-AG-1 & clip2025 & RAG & 0.3300 & 0.5700 \\
wingii-v3-gpt & WING-II & AG & 0.3100 & 0.4500 \\
ag-v2-gpt & cfdalab & AG & 0.3100 & 0.5800 \\
rag-v2-gpt & cfdalab & RAG & 0.3100 & 0.5700 \\
IDACCSabstrct-gpt4-1 & IDACCS & AG & 0.3000 & 0.5600 \\
rag-v1-gpt & cfdalab & RAG & 0.3000 & 0.5300 \\
nuggetizer & WaterlooClarke & RAG & 0.3000 & 0.5400 \\
sub\_query\_entities & clip2025 & RAG & 0.3000 & 0.5400 \\
grilllab-gpt45-gen & grilllab & RAG & 0.3000 & 0.5200 \\
IDACCS-nugg-gpt-4o & IDACCS & AG & 0.2900 & 0.4800 \\
ag-v1-gpt & cfdalab & AG & 0.2900 & 0.5500 \\
ori\_query\_entities & clip2025 & RAG & 0.2900 & 0.5300 \\
grilllab-agentic-gpt4-generation & grilllab & RAG & 0.2900 & 0.5000 \\
extractive\_rag & hltcoe-multiagt & RAG & 0.2900 & 0.5100 \\
IDACCS-hybrid-gpt4o & IDACCS & AG & 0.2800 & 0.4600 \\
wingii-3-rl-refined & WING-II & AG & 0.2800 & 0.4100 \\
uema2lab\_rag\_fewdoc & tus & RAG & 0.2800 & 0.4200 \\
hltcoe-lg.qwen & hltcoe-rerank & RAG & 0.2800 & 0.4800 \\
lg\_nt\_q4d12l3\_c & hltcoe-multiagt & RAG & 0.2800 & 0.4500 \\
IDACCS-hybrid-gpt4-1 & IDACCS & AG & 0.2700 & 0.4600 \\
ag-v2-llama & cfdalab & AG & 0.2700 & 0.5100 \\
lg\_nt\_q4d12l3 & hltcoe-multiagt & RAG & 0.2700 & 0.4800 \\
hltcoe-lg.fsrrf & hltcoe-rerank & RAG & 0.2700 & 0.5100 \\
swarm & hltcoe-multiagt & RAG & 0.2700 & 0.4900 \\
swarm\_c & hltcoe-multiagt & RAG & 0.2600 & 0.4400 \\
gptr\_e2\_q4d4 & hltcoe-multiagt & RAG & 0.2600 & 0.4900 \\
hltcoe-lg.jina & hltcoe-rerank & RAG & 0.2600 & 0.4600 \\
gptr.nt\_q4d4 & hltcoe-multiagt & RAG & 0.2600 & 0.4700 \\
gptr\_nt\_q3d3 & hltcoe-multiagt & RAG & 0.2600 & 0.4500 \\
cru-ansR-bareconf & HLTCOE & AG & 0.2500 & 0.4400 \\
cru-ansR-conf & HLTCOE & AG & 0.2400 & 0.4400 \\
cru-ansR & HLTCOE & AG & 0.2400 & 0.4300 \\
hltcoe-lg.searcher & hltcoe-rerank & RAG & 0.2400 & 0.4600 \\
cru-ablR & HLTCOE & AG & 0.2300 & 0.4000 \\
gptr\_e2\_q3d3 & hltcoe-multiagt & RAG & 0.2300 & 0.4200 \\
genSubQ\_merge & uogTr & RAG & 0.2300 & 0.4000 \\
rag-v2-llama & cfdalab & RAG & 0.2300 & 0.3900 \\
cru-ablR-conf & HLTCOE & AG & 0.2100 & 0.4000 \\
NITA\_AG\_JH & NITATREC & AG & 0.1800 & 0.3100 \\
no-llm & WING-II & AG & 0.1700 & 0.2800 \\
ag-run-1-JH & NITATREC & RAG & 0.1700 & 0.3100 \\
e5\_monot5\_searchR1 & uogTr & RAG & 0.1500 & 0.3200 \\
no-llm-refined & WING-II & AG & 0.1400 & 0.2400 \\
strd\_roll\_segment & IRIT-ISIR-EV & RAG & 0.1400 & 0.2100 \\
bm25-rz7b-2025a & ii\_research & RAG & 0.1300 & 0.2300 \\
standard\_roll & IRIT-ISIR-EV & RAG & 0.1200 & 0.1900 \\
\hline
\end{tabular}}
\vspace{0.3cm}
\caption{Summary of the RAG and AG task scores with AutoNuggetizer for 22 narratives using automatic relevance assessments.}
\label{tab:res_auto_auto_qrel}
\end{table}

\subsection{AG and RAG Task}
For the TREC 2025 RAG Track, NIST received 51 runs from 16 groups for the RAG task and 25 runs from 9 groups for the AG task.

\paragraph{Response Evaluation.}
Table~\ref{tab:res_manual_assign} presents scores for AG and RAG runs with priority levels higher than 3 under the fully manual setting (post-edited nuggets with manual assignment).
Table~\ref{tab:res_pe_auto} shows scores for AG and RAG runs under the post-edited nuggets with automatic assignment using manual relevance judgments.
Table~\ref{tab:res_auto} shows scores for AG and RAG runs under the automated setting evaluated using AutoNuggetizer using manual relevance judgments.
Table~\ref{tab:res_auto_auto_qrel} shows the fully automated setting, which also involved automatically generated relevance judgments for document selection.
Figure~\ref{fig:nugget_analysis1} showcases the correlation analysis performed for the AG and RAG tasks, analogous to the retrieval task analysis for response evaluation.
The three panels in Figure~\ref{fig:nugget_analysis1} compare the fully manual setting, Post‑Edited Nuggets + ManualAssign, with three alternative configurations: a fully automated pipeline using AutoNuggetizer with automatically generated qrels (top row), AutoNuggetizer with NIST qrels (middle row), and Post‑Edited Nuggets with NIST assessors with AutoAssign (bottom row).

In row (a) of Figure~\ref{fig:nugget_analysis1}, the strict vital score exhibits the greatest degree of scatter, indicating the weakest correlation and the lowest agreement between the fully manual and fully automated configurations. 
In row (b) of Figure~\ref{fig:nugget_analysis1}, where manually created relevance judgments are retained but nugget generation is automated, the strict vital plots show a noticeable reduction in scatter. 
This improvement reflects stronger alignment and highlights the stabilizing effect of manual relevance assessments. 
In row (c) of Figure~\ref{fig:nugget_analysis1}, which isolates the effect of nugget assignment by comparing manual and automated assignment while keeping all other components fixed, the strict vital score exhibits the highest level of correlation, with tightly clustered points and the strongest Kendall $\tau$ values observed across the three comparisons.

For the sub‑narrative coverage metric, the pattern differs somewhat. 
Rows (a) and (b) of Figure~\ref{fig:nugget_analysis1} display comparable levels of run‑level correlation, although the condition using NIST qrels shows slightly stronger agreement, suggesting that manual relevance judgments help but exert a more limited influence on this metric. 
In row (c) of Figure~\ref{fig:nugget_analysis1}, the overall scatter of individual narrative‑run points is reduced; however, the run‑level aggregated correlations decline relative to the middle row. 
This indicates that although automated assignment introduces less point‑level variability, it can still alter the ordering of runs in ways that reduce rank correlation for sub‑narrative coverage. 
Overall, the figure demonstrates that strict vital scores are more sensitive to upstream evaluation components, while sub‑narrative coverage remains comparatively stable except when nugget assignment is modified.

\begin{figure}[bth]
        \centering
        \includegraphics[width=0.45\textwidth]{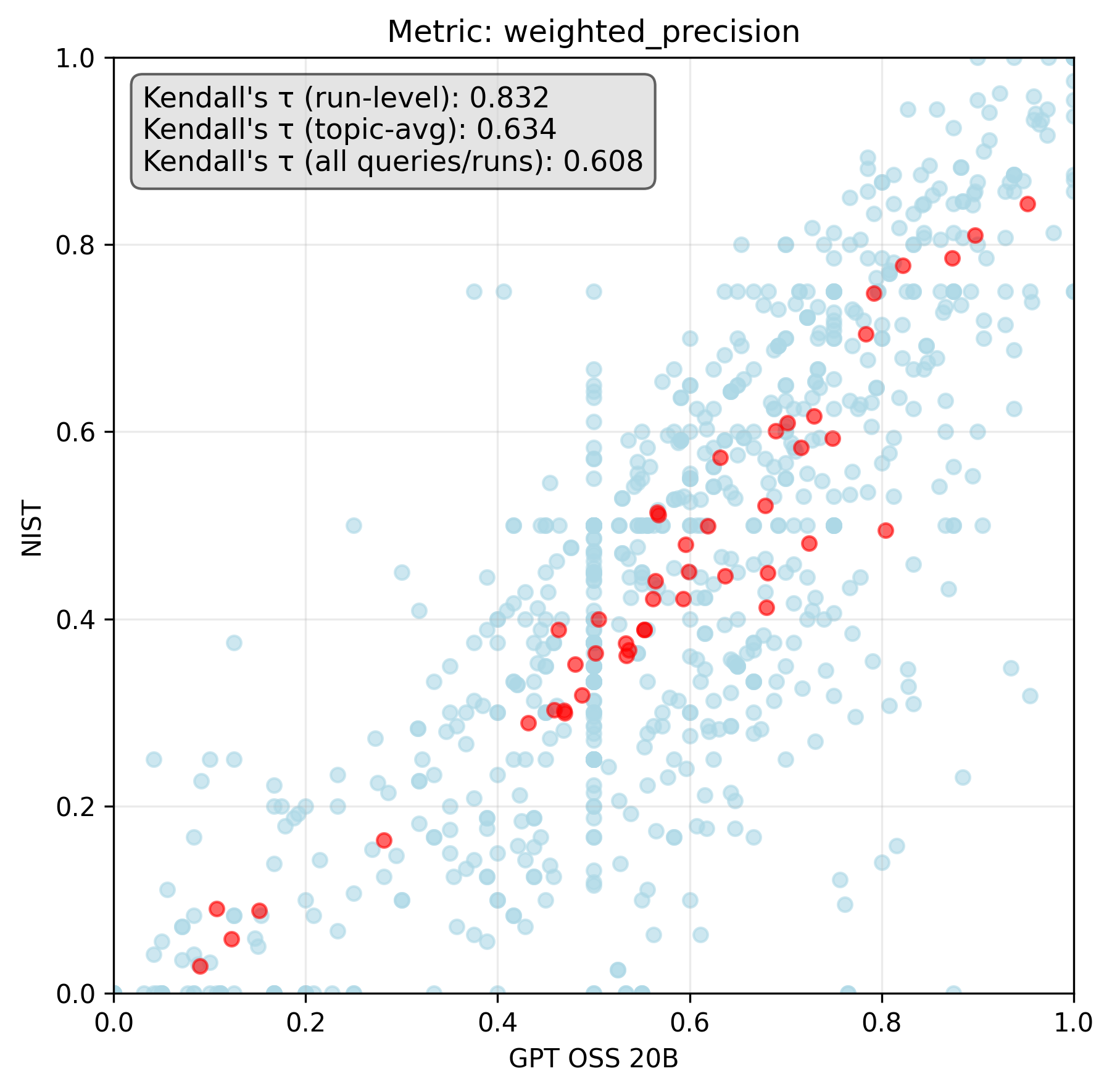}
        \includegraphics[width=0.45\textwidth]{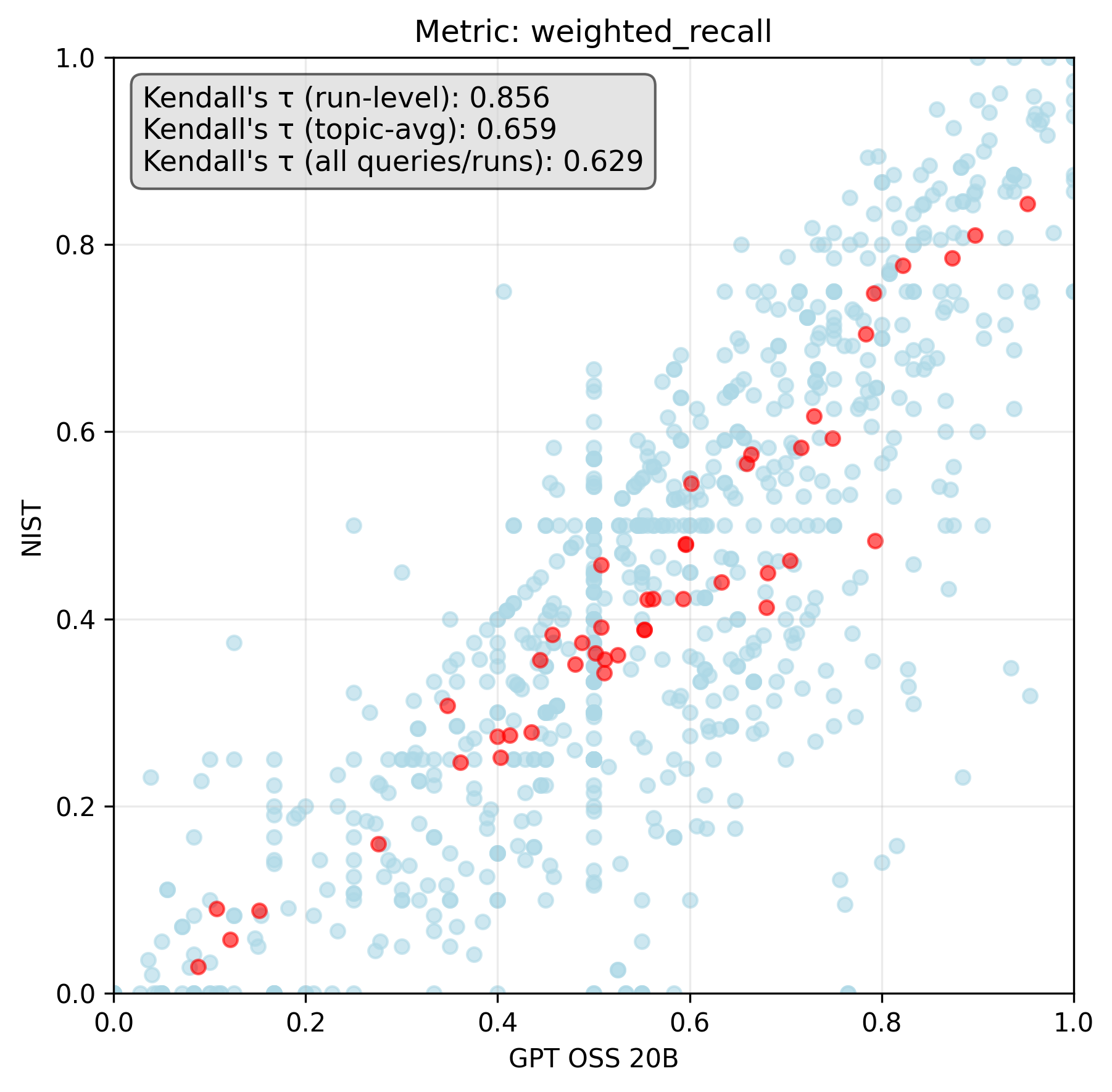}
    \caption{Comparisons between support evaluation scores calculated using GPT-OSS 120B and manual assessments.}
    \label{fig:support_analysis}
\end{figure}

\paragraph{Support Evaluation.}
Table~\ref{tab:support_nist} shows support evaluation scores for AG and RAG runs using assessments performed by NIST assessors.
Table~\ref{tab:support_gpt_oss} shows support evaluation scores for AG and RAG runs using GPT-OSS 120B as the assessor.
Figure~\ref{fig:support_analysis} compares weighted recall and weighted precision scores produced by GPT‑OSS 120B in the automatic setting against the manual setting, produced by NIST assessors. 
For both metrics, the scatterplots show a clear positive relationship, with high run‑level Kendall’s $\tau$ values indicating strong overall alignment between the model‑based and manual assessments. 
Although some dispersion appears at the individual narrative level, the aggregated red points fall close to the diagonal, demonstrating that GPT‑OSS 120B captures the overall ranking structure of systems reasonably well. 
The similarity in patterns across both weighted recall and weighted precision suggests that GPT‑OSS 120B provides a stable approximation of human scoring for metrics that integrate both relevance and nugget importance. 
Overall, Figure~\ref{fig:support_analysis} indicates that while fine‑grained variability persists, GPT‑OSS 120B serves as a reliable surrogate for manual evaluation at the run level.

\begin{table}[t]
\centering
\scalebox{0.65}{
\begin{tabular}{l l l cc}
\hline
Run ID & Group & Task & weighted precision & weighted recall \\ \hline
no-llm-refined & WING-II & AG & 0.8435 & 0.8435 \\
cru-ablR & HLTCOE & AG & 0.8098 & 0.8098 \\
IDACCS-hybrid-gpt4-1 & IDACCS & AG & 0.7857 & 0.7857 \\
nugget-generation & GenAIus & AG & 0.7780 & 0.7780 \\
bm25-rz7b-2025a & ii\_research & RAG & 0.7792 & 0.7635 \\nick
cluster-generation & GenAIus & AG & 0.7483 & 0.7483 \\
cru-ansR & HLTCOE & AG & 0.7044 & 0.7044 \\
wingii-3-rl-refined & WING-II & AG & 0.6903 & 0.6903 \\
Rerank-Top50\_v2 & clip2025 & AG & 0.6171 & 0.6171 \\
cru-ansR-conf & HLTCOE & AG & 0.5927 & 0.5927 \\
Rerank-Top50\_v3 & clip2025 & AG & 0.5831 & 0.5831 \\
wingii-v3-gpt & WING-II & AG & 0.5778 & 0.5778 \\
gptr.nt\_q4d4 & hltcoe-multiagt & RAG & 0.6007 & 0.5758 \\
gptr\_e2\_q4d4 & hltcoe-multiagt & RAG & 0.6099 & 0.5662 \\
gpt41 & UTokyo & AG & 0.5725 & 0.5446 \\
selector-agent-trim & ncsu-las & RAG & 0.4870 & 0.4870 \\
lg\_nt\_q4d12l3\_c & hltcoe-multiagt & RAG & 0.4947 & 0.4838 \\
r\_4method\_ag\_gpt41 & UTokyo & RAG & 0.4995 & 0.4801 \\
IDACCS-nugg-gpt-4-1 & IDACCS & AG & 0.4797 & 0.4797 \\
hltcoe-lg.searcher & hltcoe-rerank & RAG & 0.4810 & 0.4622 \\
standard\_roll & IRIT-ISIR-EV & RAG & 0.5000 & 0.4615 \\
rag25\_qwen3\_20\_ag & coordinators & AG & 0.5110 & 0.4576 \\
auto\_plan & WaterlooClarke & RAG & 0.4493 & 0.4493 \\
hltcoe-lg.qwen & hltcoe-rerank & RAG & 0.4458 & 0.4392 \\
auto\_selected & WaterlooClarke & RAG & 0.4213 & 0.4213 \\
combined & WaterlooClarke & RAG & 0.4212 & 0.4212 \\
r\_2method\_ag\_gpt41 & UTokyo & RAG & 0.4504 & 0.4208 \\
hltcoe-gpt5.searcher & hltcoe-rerank & RAG & 0.4124 & 0.4124 \\
grilllab-agentic-gpt4-generation & grilllab & RAG & 0.5211 & 0.3911 \\
NITA\_AG\_JH & NITATREC & AG & 0.3888 & 0.3888 \\
ag-run-1-JH & NITATREC & RAG & 0.3888 & 0.3888 \\
no-reranker & MITLL & RAG & 0.3887 & 0.3831 \\
LAS-agentic-RAG-selector & ncsu-las & RAG & 0.4410 & 0.3745 \\
LAS-agentic-RAG-agent & ncsu-las & RAG & 0.3697 & 0.3697 \\
Kun-Third & RMIT-IR & RAG & 0.3635 & 0.3635 \\
no-decomp & MITLL & RAG & 0.3671 & 0.3615 \\
full & MITLL & RAG & 0.3738 & 0.3567 \\
strd\_roll\_segment & IRIT-ISIR-EV & RAG & 0.4067 & 0.3532 \\
IDACCSabstrct-gpt4-1 & IDACCS & AG & 0.3514 & 0.3514 \\
uema2lab\_B4 & tus & RAG & 0.3606 & 0.3424 \\
grilllab-gpt45-gen & grilllab & RAG & 0.5137 & 0.3071 \\
rag-v1-gpt & cfdalab & RAG & 0.3024 & 0.2794 \\
rag-v2-gpt & cfdalab & RAG & 0.3190 & 0.2756 \\
ag-v1-gpt & cfdalab & AG & 0.2892 & 0.2742 \\
rag25\_qwen3\_50\_ag & coordinators & AG & 0.3393 & 0.2725 \\
ag-v2-gpt & cfdalab & AG & 0.2997 & 0.2518 \\
ag-v2-llama & cfdalab & AG & 0.3031 & 0.2465 \\
rag-v2-llama & cfdalab & RAG & 0.2898 & 0.2225 \\
KG-AG-1 & clip2025 & RAG & 0.1637 & 0.1599 \\
uema2lab\_rag\_fewdoc & tus & RAG & 0.1074 & 0.1074 \\
e5\_monot5\_searchR1 & uogTr & RAG & 0.0907 & 0.0907 \\
sub\_query\_entities & clip2025 & RAG & 0.0883 & 0.0883 \\
uema2lab\_base & tus & RAG & 0.0673 & 0.0673 \\
ori\_query\_entities & clip2025 & RAG & 0.0582 & 0.0573 \\
genSubQ\_merge & uogTr & RAG & 0.0294 & 0.0285 \\
\hline
\end{tabular}}
\vspace{0.3cm}
\caption{Summary of the support evaluation for RAG and AG task for 17 narratives.}
\label{tab:support_nist}
\end{table}

\begin{table}[t]
\centering
\scalebox{0.65}{
\begin{tabular}{l l l cc}
\hline
Run ID & Group & Task & weighted precision & weighted recall \\ \hline
no-llm-refined & WING-II & AG & 0.9522 & 0.9522 \\
bm25-rz7b-2025a & ii\_research & RAG & 0.9298 & 0.9189 \\
cru-ablR & HLTCOE & AG & 0.8972 & 0.8972 \\
IDACCS-hybrid-gpt4-1 & IDACCS & AG & 0.8734 & 0.8734 \\
nugget-generation & GenAIus & AG & 0.8220 & 0.8220 \\
lg\_nt\_q4d12l3\_c & hltcoe-multiagt & RAG & 0.8039 & 0.7930 \\
cluster-generation & GenAIus & AG & 0.7920 & 0.7920 \\
cru-ansR & HLTCOE & AG & 0.7835 & 0.7835 \\
wingii-v3-gpt & WING-II & AG & 0.7833 & 0.7833 \\
wingii-3-rl-refined & WING-II & AG & 0.7722 & 0.7722 \\
cru-ansR-conf & HLTCOE & AG & 0.7487 & 0.7487 \\
Rerank-Top50\_v2 & clip2025 & AG & 0.7293 & 0.7293 \\
Rerank-Top50\_v3 & clip2025 & AG & 0.7158 & 0.7158 \\
hltcoe-lg.searcher & hltcoe-rerank & RAG & 0.7242 & 0.7043 \\
auto\_plan & WaterlooClarke & RAG & 0.6810 & 0.6810 \\
hltcoe-gpt5.searcher & hltcoe-rerank & RAG & 0.6799 & 0.6799 \\
gptr.nt\_q4d4 & hltcoe-multiagt & RAG & 0.6893 & 0.6640 \\
gptr\_e2\_q4d4 & hltcoe-multiagt & RAG & 0.7021 & 0.6597 \\
hltcoe-lg.qwen & hltcoe-rerank & RAG & 0.6366 & 0.6331 \\
gpt41 & UTokyo & AG & 0.6320 & 0.6019 \\
r\_4method\_ag\_gpt41 & UTokyo & RAG & 0.6186 & 0.5957 \\
IDACCS-nugg-gpt-4-1 & IDACCS & AG & 0.5956 & 0.5956 \\
auto\_selected & WaterlooClarke & RAG & 0.5935 & 0.5935 \\
LAS-agentic-RAG-agent & ncsu-las & RAG & 0.5773 & 0.5773 \\
selector-agent-trim & ncsu-las & RAG & 0.5669 & 0.5669 \\
combined & WaterlooClarke & RAG & 0.5619 & 0.5619 \\
standard\_roll & IRIT-ISIR-EV & RAG & 0.6154 & 0.5577 \\
r\_2method\_ag\_gpt41 & UTokyo & RAG & 0.5988 & 0.5558 \\
NITA\_AG\_JH & NITATREC & AG & 0.5527 & 0.5527 \\
ag-run-1-JH & NITATREC & RAG & 0.5527 & 0.5527 \\
strd\_roll\_segment & IRIT-ISIR-EV & RAG & 0.6091 & 0.5377 \\
no-decomp & MITLL & RAG & 0.5364 & 0.5248 \\
full & MITLL & RAG & 0.5333 & 0.5115 \\
uema2lab\_B4 & tus & RAG & 0.5341 & 0.5107 \\
rag25\_qwen3\_20\_ag & coordinators & AG & 0.5675 & 0.5079 \\
grilllab-agentic-gpt4-generation & grilllab & RAG & 0.6784 & 0.5076 \\
Kun-Third & RMIT-IR & RAG & 0.5022 & 0.5022 \\
LAS-agentic-RAG-selector & ncsu-las & RAG & 0.5641 & 0.4881 \\
IDACCSabstrct-gpt4-1 & IDACCS & AG & 0.4807 & 0.4807 \\
rag25\_test\_qwen3\_20 & coordinators & RAG & 0.5464 & 0.4623 \\
no-reranker & MITLL & RAG & 0.4632 & 0.4569 \\
rag25\_test\_qwen3\_50 & coordinators & RAG & 0.5050 & 0.4442 \\
rag-v1-gpt & cfdalab & RAG & 0.4692 & 0.4353 \\
rag-v2-gpt & cfdalab & RAG & 0.4881 & 0.4126 \\
ag-v2-gpt & cfdalab & AG & 0.4695 & 0.4029 \\
ag-v1-gpt & cfdalab & AG & 0.4318 & 0.4000 \\
rag-v2-llama & cfdalab & RAG & 0.4906 & 0.3793 \\
rag25\_qwen3\_50\_ag & coordinators & AG & 0.4684 & 0.3730 \\
ag-v2-llama & cfdalab & AG & 0.4590 & 0.3614 \\
grilllab-gpt45-gen & grilllab & RAG & 0.5662 & 0.3475 \\
uema2lab\_rag\_fewdoc & tus & RAG & 0.2849 & 0.2849 \\
KG-AG-1 & clip2025 & RAG & 0.2813 & 0.2754 \\
sub\_query\_entities & clip2025 & RAG & 0.1515 & 0.1515 \\
ori\_query\_entities & clip2025 & RAG & 0.1224 & 0.1211 \\
uema2lab\_base & tus & RAG & 0.1171 & 0.1171 \\
e5\_monot5\_searchR1 & uogTr & RAG & 0.1074 & 0.1074 \\
genSubQ\_merge & uogTr & RAG & 0.0902 & 0.0878 \\
\hline
\end{tabular}}
\vspace{0.3cm}
\caption{Automatic support evaluation (weighted precision and recall) for RAG and AG tasks with GPT-OSS 120B as the assessor for 17 narratives.}
\label{tab:support_gpt_oss}
\end{table}

\subsection{Relevance Judgment (RJ) Task}
For the TREC 2025 RAG Track, NIST received 36 runs from 5 groups for the Relevance Judgment task.
Table~\ref{tab:qrels} shows alignment evaluation when the participant's submissions are compared against human judgment.
The best runs achieve agreement fractions between 0.30--0.34, suggesting that automated relevance judgment remains a challenging task with limited agreement relative to human assessors, even among the best-performing runs.

\begin{table}[t]
\centering
\scalebox{0.65}{
\begin{tabular}{l l l cccc}
\hline
RUN ID & Group & Run Type & agreement\_frac & kappa\_val \\ \hline
gpt-oss-120b-sn-med & coordinators & automatic & 0.3400 & 0.2100 \\
jcru-ablR & HLTCOE & automatic & 0.3200 & 0.0100 \\
jcru-ansR-all & HLTCOE & automatic & 0.3200 & N/A \\
jcru-ablR-all & HLTCOE & automatic & 0.3200 & N/A \\
jcru-ansR & HLTCOE & automatic & 0.3200 & 0.0000 \\
gpt-oss-120b-sn-low & coordinators & automatic & 0.3100 & 0.1700 \\
gpt-oss-20b-high & coordinators & automatic & 0.3100 & 0.1200 \\
gpt-oss-120b-sn-high & coordinators & automatic & 0.3100 & 0.1200 \\
gpt-oss-120b-high & coordinators & automatic & 0.3000 & 0.1400 \\
gpt-oss-120b-med & coordinators & automatic & 0.3000 & 0.1600 \\
gpt-oss-20b-sn-low & coordinators & automatic & 0.3000 & 0.1000 \\
gpt\_5-sub-narr & coordinators & automatic & 0.2900 & 0.1000 \\
gpt-oss-120b-low & coordinators & automatic & 0.2900 & 0.1300 \\
Qwen3-30BThink-sn & coordinators & automatic & 0.2900 & 0.1300 \\
gpt-oss-20b-medium & coordinators & automatic & 0.2900 & 0.1000 \\
gpt-oss-20b-sn-high & coordinators & automatic & 0.2900 & 0.0900 \\
gemini\_2\_5\_pro & coordinators & automatic & 0.2900 & 0.1500 \\
gpt-oss-20b-sn-med & coordinators & automatic & 0.2800 & 0.0800 \\
unique\_cluster\_cnt & GenAIus & automatic & 0.2800 & 0.0700 \\
cluster\_cnt & GenAIus & automatic & 0.2800 & 0.0900 \\
nugget\_cnt & GenAIus & automatic & 0.2800 & 0.1300 \\
citation\_cnt & GenAIus & automatic & 0.2800 & 0.0700 \\
gpt-oss-20b-low & coordinators & automatic & 0.2800 & 0.0900 \\
ensemble\_umbrela1 & coordinators & automatic & 0.2700 & 0.1800 \\
norm\_nugget\_cnt & GenAIus & automatic & 0.2600 & 0.0600 \\
Qwen3-30B-Think & coordinators & automatic & 0.2600 & 0.1200 \\
duth\_hybrid\_qwen\_cal & DUTH\_XANTHI & automatic & 0.2300 & 0.0400 \\
duth\_hybrid\_qwencon & DUTH\_XANTHI & automatic & 0.2300 & 0.0500 \\
hybrid\_stable\_loose2 & DUTH\_XANTHI & automatic & 0.2200 & -0.0100 \\
duth\_hybrid\_stableri & DUTH\_XANTHI & automatic & 0.2200 & -0.0100 \\
Qwen3-30BInstruct-sn & coordinators & automatic & 0.2200 & 0.0500 \\
Qwen3-30B-Instruct & coordinators & automatic & 0.2200 & 0.1100 \\
gpt\_4\_1-sub-narr-2 & coordinators & automatic & 0.2100 & 0.1900 \\
NITA-Qrels & NIT Agartala & automatic & 0.1900 & 0.0600 \\
gpt\_4-1-sub-narr & coordinators & automatic & 0.1800 & 0.1400 \\
gpt\_4-1 & coordinators & automatic & 0.1800 & 0.1500 \\
duth\_stablelm2\_rj\_v1 & DUTH\_XANTHI & manual & 0.0200 & 0.0000 \\
\hline
\end{tabular}}
\vspace{0.3cm}
\caption{Alignment comparison for relevance judgment task for 22 narratives.}
\label{tab:qrels}
\end{table}

\section{Conclusion}
The \trecrag Track represents a significant step toward advancing RAG systems capable of handling complex, long context narratives. 
By shifting the focus from keyword-based queries to nuanced, multi-sentence information needs this year, the track received over 150 submissions that integrated deep reasoning, evidence synthesis, and transparent attribution. 
The multi-layered evaluation framework ensures rigorous assessment of retrieval quality, factual grounding, and response completeness, setting a high standard for future research. 
Through these innovations, the track seeks to inspire the development of robust, trustworthy AI systems that can meet the demands of real-world information seeking and contribute to the broader progress in IR and NLP.

\section*{Acknowledgements}

This research was supported in part by the Natural Sciences and Engineering Research Council (NSERC) of Canada and by an Institute of Information.
We extend our sincere appreciation to all participants who submitted work to the 2025 track. 
We are also grateful to Ian Soboroff and Hoa Dang at NIST for coordinating the manual assessment process, as well as to all assessors whose expertise was essential to the evaluation.

\clearpage
\bibliography{main}
\bibliographystyle{acl_natbib}
\clearpage


\appendix



\end{document}